\documentclass[aps,preprint,superscriptaddress,longbibliography,prl]{revtex4-2}
\usepackage{graphicx, amsmath, verbatim, dsfont, amsfonts, color, amssymb, comment}
\usepackage{hyperref}
\usepackage{braket}
\usepackage[us]{datetime}
\usepackage{comment}

\graphicspath{{figures/}}
\newcommand*{\bp}{$B_\perp$}

\newcommand*{\di}{$\Delta I$}
\newcommand*{\isp}{$\langle I_{sw}^+\rangle$}
\newcommand*{\isn}{$\langle I_{sw}^-\rangle$}
\newcommand*{\isw}{$\langle I_{sw}\rangle$}

\begin{document}
\title{Diamagnetic mechanism of critical current non-reciprocity in multilayered superconductors}
\author{Ananthesh~Sundaresh}
\affiliation{Department of Physics and Astronomy, Purdue University, West Lafayette, IN 47907 USA}
\author{Jukka I.~V\"ayrynen}
\affiliation{Department of Physics and Astronomy, Purdue University, West Lafayette, IN 47907 USA}
\author{Yuli~Lyanda-Geller}
\affiliation{Department of Physics and Astronomy, Purdue University, West Lafayette, IN 47907 USA}
\author{Leonid~P.~Rokhinson}
\email{leonid@purdue.edu}
\affiliation{Department of Physics and Astronomy, Purdue University, West Lafayette, IN 47907 USA}
\affiliation{Department of Electrical and Computer Engineering, Purdue University, West Lafayette, IN 47907 USA}

\date{\today}

\maketitle

\textbf{Recent excitement in observation of non-reciprocal critical current (NRC) is motivated by a suggestion that ``superconducting diode effect'' may be an intrinsic property of non-centrosymmetric superconductors with strong spin-orbit interactions\cite{Ando2020}. Theoretically it has been understood that linear in the Cooper pair momentum terms, caused by the Rashba spin-orbit and Zeemann interactions or, more generally, any symmetry-allowed Lifshitz invariants\cite{Agterberg2012} in uniform singlet superconductors, do not contribute to the supercurrent, although the role of higher-order terms remains unclear\cite{Daido2022,He2022}. In this work we show that critical current non-reciprocity is a generic property of multilayered superconductor structures in the presence of magnetic field-generated diamagnetic currents. In the regime of an intermediate coupling between the layers, the Josephson vortices are predicted to form at high fields and currents. We report the observation of NRC in nanowires fabricated from InAs/Al heterostructures. The effect is independent of the crystallographic orientation of the wire, ruling out an intrinsic origin of NRC. Non-monotonic NRC evolution with magnetic field is consistent with the generation of diamagnetic currents and formation of the Josephson vortices. This extrinsic NRC mechanism can be used to design novel devices for superconducting circuits.}

Diodes are the most basic elements of semiconductor electronics and development of superconducting diodes can extend the functionality of the superconducting circuitry. A non-reciprocal critical current (NRC) in a {\it multiply-connected} superconductors is a well known effect and can be readily observed in, e.g., an asymmetric superconducting rings~\cite{Burlakov2014}. An implicit suggestion that NRC may be an \textit{intrinsic} property of non-centrosymmetric superconductors~\cite{Ando2020} generated a renewed theoretical and experimental interest motivated by an analogy with the non-reciprocal resistivity due to the magnetochiral effect, which can appear in uniform materials with broken spatial and time-reversal symmetry~\cite{Rikken2005}.  However, a direct analogy between corrections to resistivity and superconducting current is misleading because the anisotropy of scattering is caused by the spin-orbit effects, while the proposed origin of non-reciprocity in singlet-pairing superconductors is a spin-independent Lifshitz invariant~\cite{Levitov1985,Edelstein1996}. It has been demonstrated in the literature \cite{Agterberg2012} that in uniform singlet superconductors in constant magnetic field the Lifshitz invariants can be eliminated by a gauge (Galilean) transformation from both the Ginzburg-Landau (GL) equation and the expression for the supercurrent, so that linear in Cooper pair momentum terms do not lead to non-reciprocity. Phenomenological treatment shows that cubic in the Cooper pair momentum terms can lead to NRC corrections~\cite{Daido2022}. It has been also suggested that the Rashba terms in the electron spectrum contribute to non-reciprocity within the formalism of quasiclassical Eilenberger equations \cite{Ilic2022}. A symmetry analysis and microscopic calculations of the Cooperon propagator in the presence of the Zeemann effect  and a linear or cubic Dresselhaus spin-orbit interactions in the electron spectrum show that the NRC magnitude and sign depend on the crystallographic orientation of the supercurrent flow~\cite{Lyanda-Geller2022}. Such anisotropy should characterize non-reciprocity in  superconductor/semiconductor heterostructures and cubic uniform singlet noncentrosymmetric superconductors. Another suggested mechanism of NRC is the formation of non-uniform currents in superconducting multilayers \cite{Vodolazov2018,Zinkl2021}.

In this paper we show that NRC naturally arises in the presence of the magnetic-field--generated diamagnetic currents when neighboring layers in multilayer superconductors are strongly coupled. The total current through a multilayer structure is divided between the layers as an inverse ratio of their kinetic inductances. Initially only one of the layers reaches the maximum (critical) current as the total current increases. A further current increase in the superconducting state requires generating a phase difference between the layers, which adds Josephson energy penalty to the total energy, and in the case of strong interlayer coupling (large Josephson energy) the whole system transitions into a normal state. Field-generated diamagnetic currents will either increase or decrease an external current where transition to the normal state occurs, thus leading to the NRC. In the regime of intermediate interlayer coupling strengths, an interlayer phase difference can change by $2\pi$, leading to the formation of Josephson vortices. Experimentally, we report observation of NRC in nanowires fabricated from InAs/Al heterostructures. The observed non-monotonic evolution of NRC with magnetic field is consistent with the formation of Josephson vortices. Our findings show that the extrinsic contribution to NRC is generic to multilayer superconductors, and may provide a compelling explanation to the NRC observed in Refs. ~\onlinecite{Ando2020} and \onlinecite{Narita2022}, in the latter work magnetic flux produced by a Cr layer generates opposite diamagnetic currents in the adjacent Nb and V layers.

The term ``superconducting diode effect'' has been used to describe NRC in different systems, including thin superconducting films~\cite{Bauriedl2022,Jiang2022,Lyu2021,Suri2022,Ustavschikov2022,Hou2022} and Josephson junctions~\cite{Diez-Merida2021,Baumgartner2022,Pal2022,Jeon2022,Gupta2022,Golod2022,Wu2022,John2022,Bianca2022}. In the former experiments the presence of out-of-plane magnetic field and formation of vortices is essential for the observation of NRC, in this case the critical current is determined by the strength and symmetry of the flux pinning potential. In the latter case the critical current in Josephson junctions is determined by the overlap of Andreev states. In this paper we restrict our discussion to the origin of NRC in long nanowires, where critical current is determined by the de-pairing velocity of Cooper pairs (the Bardeen limit \cite{Bardeen1962}).

\begin{figure}[t]
\centering
\includegraphics[width=0.9\textwidth]{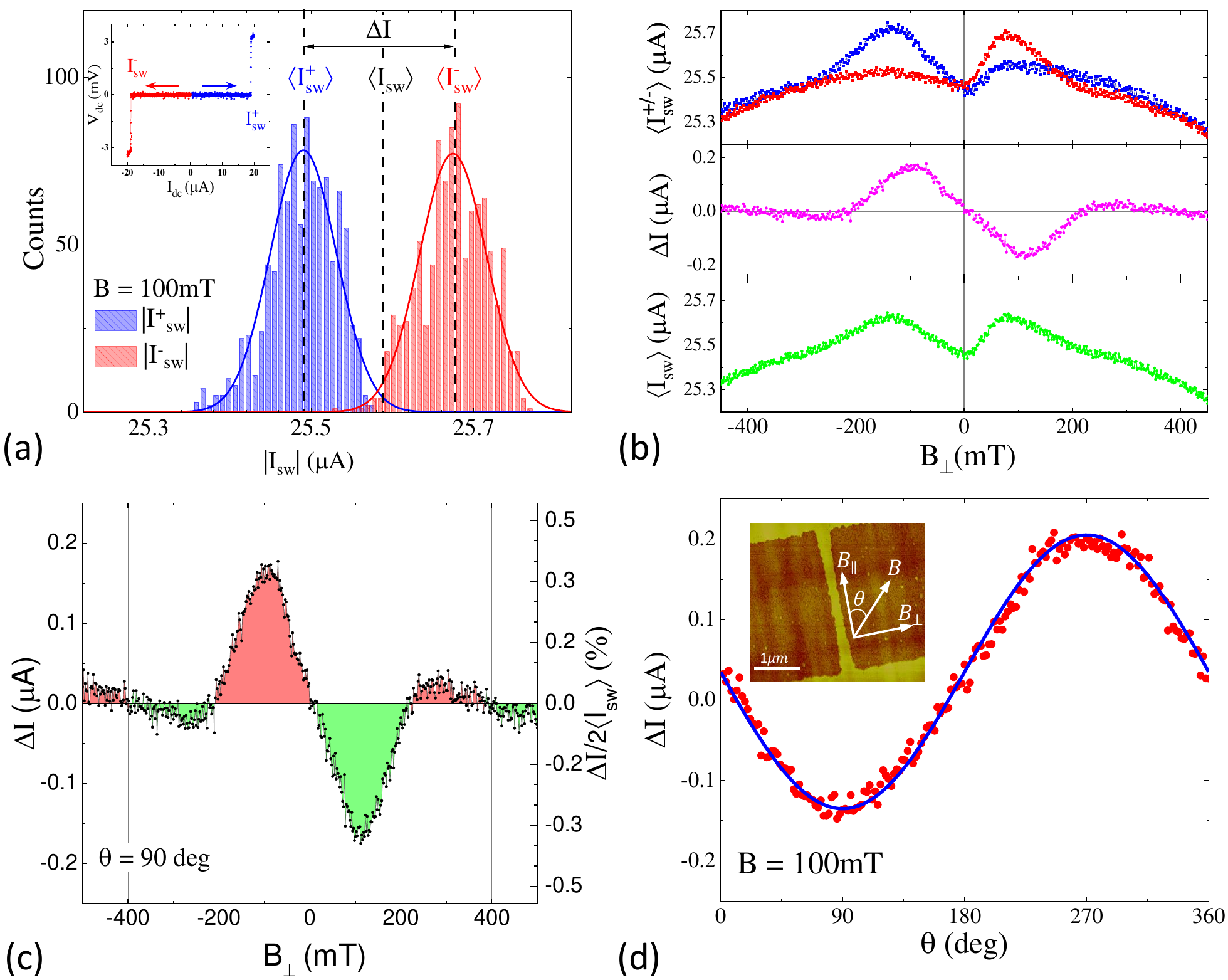}
\caption{\textbf{Non-reciprocal critical current in Al/InAs nanowires.} (a) Histograms of switching currents for 10,000 positive $I_{sw}^+$ and negative $I_{sw}^-$ current sweeps performed at $T=30$ mK and $B_\bot=100$ mT. Inset shows a typical current-voltage characteristic. (b) Average switching current for positive \isp\ and negative \isn\ sweeps, non-reciprocal difference $\Delta I=\langle I_{sw}^+\rangle-\langle I_{sw}^-\rangle$ and an average of all sweeps \isw\ is plotted as a function of in-plane magnetic field $B_\bot$. In (c) enlarged \di\ data is colored to signify non-monotonic field dependence and multiple sign changes. (d) Dependence of \di\ on in-plane field orientation is measured at a constant $B=100$ mT. Blue line is a fit with a sine function. Insert shows an AFM image of a $3\ \mu$m-long wire connected to wide contacts, yellow areas are Al, in darker areas Al is removed and InAs is exposed.}
\label{f1}
\end{figure}

We have studied switching currents $I_{sw}$ defining a transition from superconducting to normal state in nanowires fabricated from Al/InGaAs/InAs/InGaAs heterostructures~\cite{Shabani2016}, where patterned Al top layer forms a nanowire and induces superconductivity in a high mobility InAs quantum well via the proximity effect. An AFM micrograph of a typical device is shown in the inset in Fig.~\ref{f1}. A typical current-voltage characteristic exhibits a sharp switching transition limited by the current resolution ($<5$ nA for the fastest sweep rates used in our experiments). A histogram of switching currents $I_{sw}^{\pm}$ for positive (+) and negative (-)  current sweeps is shown in Fig.~\ref{f1}(a) for 10,000 sweeps. Field dependence of average values \isp\ and \isn\ is plotted in Fig.~\ref{f1}(b) for the in-plane field $B_\bot$ perpendicular to the wire. The \isp\ and \isn\ can be separated into a symmetric $\langle I_{sw}\rangle=(\langle I_{sw}^+\rangle+\langle I_{sw}^-\rangle)/2$ and asymmetric $\Delta I=\langle I_{sw}^+\rangle - \langle I_{sw}^-\rangle$ parts, the latter being the non-reciprocal component of the supercurrent. Both \isw\ and \di\ are non-monotonic functions of magnetic field. As shown in the Supplement, a minima of $\langle I_{sw}\rangle$ at low fields vanishes above 350 mK (0.3 $T_C$), while there is no change in $\Delta I$ at least up to 750 mK ($>0.6 T_C$). This difference in energy scales for the appearance of NRC and non-monotonic evolution of \isw\ indicates that these are unrelated phenomena, and below we focus on the origin of NRC. Some devices were fabricated with a top gate, which allows electrostatic control of the electron density in the InAs layer not covered by Al; we found that depletion of the 2D electron gas in the exposed InAs results in a slight increase of \isw\ but does not affect \di. Similar field effect has been observed previously in superconductor nanodevices~\cite{De_Simoni2018}  and was attributed to the presence of quasiparticles~\cite{Golokolenov2020}, a conclusion consistent with the observed gate dependence of the \isw.

\begin{figure}[h]
\centering
\includegraphics[width=1\textwidth]{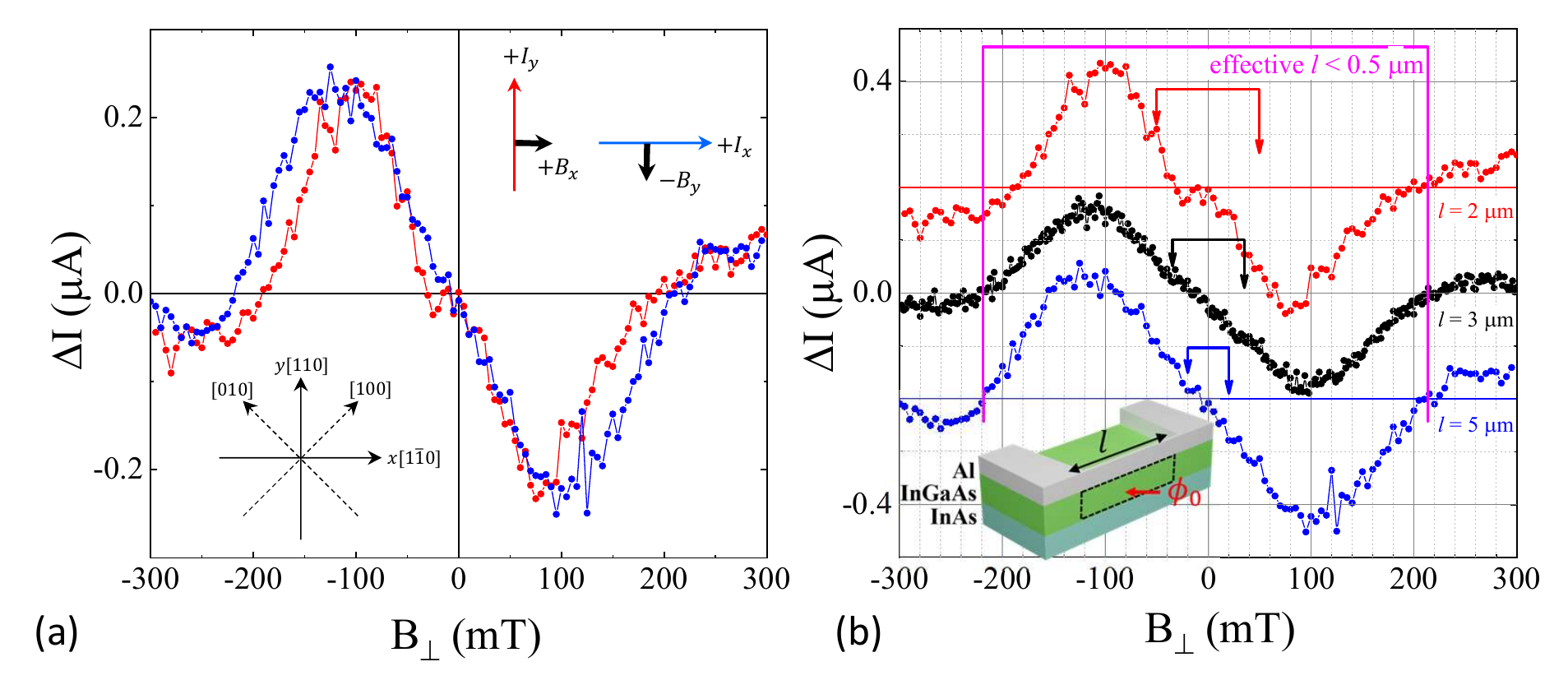}
\caption{\textbf{Dependence of NRC on the nanowire length and crystallographic orientation.} (a) NRC is plotted for two 2 $\mu$m-long  wires oriented along $[110]$ and $[1\overline{1}0]$ crystallographic axes. Insets define mutual orientataion of wires and fields. (b) NRC for 2, 3, and 5 $\mu$m wires. The top and bottom curves are shifted vertically by 0.2 $\mu$A. Brackets with arrows indicate a maximum $\Delta B$ needed to insert a flux $\phi_0=h/2e$ in the area defined by the corresponding wire lengths, as indicated by a dashed loop in the inset. An effective length for the period marked by a magenta bracket is $l=0.5\ \mu$m for the same loop. }
\label{f2}
\end{figure}

Unlike the linear in Cooper pair momentum terms, higher order terms cannot be removed by gauge transformation and it was shown that the presence of terms  $\sim\alpha_3Q^3\Delta^2$ cubic in the Cooper pair momentum in an expansion of the Ginsburg-Landau coefficients  can generally lead to non-zero \di\ which is a non-monotonic function of $B$ and can even change sign~\cite{Daido2022,Ilic2022} (here $\mathbf{Q}=-i\hbar\nabla-2e\mathbf{A}$ is a generalized Cooper pair momentum, $\mathbf{A}$ is electromagnetic vector-potential). However, for proximitized InAs layer, a generation of the terms higher order in the Cooper pair momentum in the presence of the Rashba spin-orbit and Zeeman interactions coexists with a similar generation of such terms due to the Dresselhaus spin-orbit interactions. The importance of the Dresselhaus-like terms in the electron spectrum is not limited to proximity structures, and they can play significant role in any noncentrosymmetric material. Investigation of realistic cubic terms in the Cooper pair momentum showed~\cite{Lyanda-Geller2022} that nonreciprocity  becomes highly anisotropic as a result of Dresselhaus-induced contribution. For comparison with experiments, it is instructive to express the odd in Cooper pair momentum part of the kinetic energy in coordinates rotated by $\pi/2$ with respect to the principal crystallographic axes of InAs, where $\hat{x}||[1\overline{1}0]$ and $\hat{y}||[110]$, see insert in Fig.~\ref{f2}(a). In these coordinates, the cubic in the Cooper pair momentum kinetic term originating from the cubic Dresselhaus electron spin-orbit interaction reads
\begin{equation}
    f_k=|\kappa(B_yQ^3_x + B_xQ^3_y-Q_xQ_y[B_xQ_y+B_yQ_x])\Delta |^2,
\end{equation}
where coefficient $\kappa$ contains the Dresselhaus constant $\beta_D$ and other material parameters. The resulting NRC correction to the supercurrent is
\begin{equation}
     \Delta I \propto (B_yI_x^2+B_xI_y^2).
\end{equation}
This correction is independent of the sign of $I$ and is added or subtracted to the $B=0$ current value depending on the direction of the current flow. Here $B_x$ and $B_y$ enter symmetrically for wires oriented along $x$ and $y$. However, in the configuration with the current $I \| \hat{x}$ and magnetic field $B_y$ and the configuration with $I\|\hat{y}$ and $B_x$, this expression has opposite signs for the same mutual orientation of $I$ and $B$, see inset in Fig.~\ref{f2}(a). Thus, the Dresselhaus-induced contribution results in NRC with opposite sign for wires oriented along $[1\overline{1}0]$ and $[110]$ crystallographic axis. The cubic (and generally all odd) in Cooper pair momentum terms originating from the Rashba electronic interactions, when added with the Dresselhaus-induced terms, will produce anisotropy in the absolute value of NRC, and, in particular, different values of non-reciprocal asymmetrical component of the current for those two directions. Theoretical investigation of electronic spectra of these systems \cite{Winkler2019} suggests that in narrow InAs quantum wells cubic Dresselhaus terms are larger than the Rashba terms. The lower limit for the value of the Dresselhaus contribution can  be extracted from the total spin-orbit anisotropy (which is defined by the ratio between a linear Rashba, and a linear and cubic Dresselhaus terms in electronic spectrum), which was measured to be 70\% in spin-galvanic and circular photogalvanic experiments \cite{Giglberger2007} and $>15\%$ in transport experiments\cite{Farzaneh2022,Baumgartner2022a}. Such anisotropies must result in the corresponding crystallographic anisotropy of the NRC, which is not observed in our experiments, Fig.~\ref{f2}(a). Therefore, we conclude that the NRC we observed is not intrinsic. The observed NRC does not depend on the wire length, Fig.~\ref{f2}b, which rules out trivial effects related to the formation of spurious loops due to the presence of wire/contact boundaries.

\begin{figure}[h]
\centering
\includegraphics[width=1\textwidth]{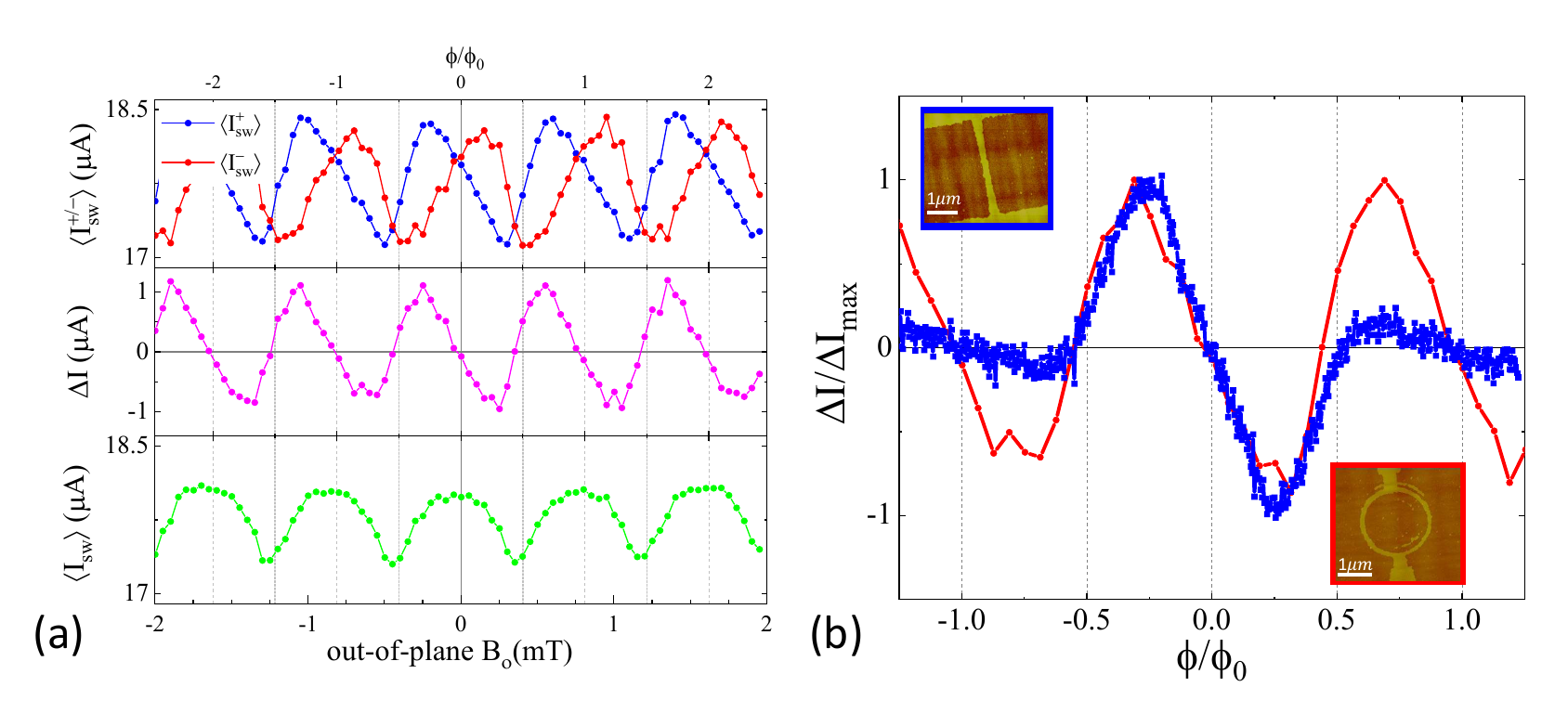}
\caption{\textbf{NRC in an asymmetric superconducting loop.} (a) An average switching current for positive \isp\ and negative \isn\ sweeps, non-reciprocal difference $\Delta I=\langle I_{sw}^+\rangle+\langle I_{sw}^-\rangle$ and an average of all sweeps \isw\ plotted as a function of out-of-plane magnetic field $B_o$ for a loop shown in the insert in (b). Note that \isw\ is maximal while $\Delta I=0$ when the flux $\phi=n\phi_0$. In (b) \di\ for the nanowire and the loop are plotted together as a function of a reduced flux $\phi/\phi_0$, where we used $S_{wire}=0.0052\ \mu$m$^2$ for the effective area in the wire and $S_{loop}=2.59\ \mu$m$^2$ in the loop.}
\label{f3}
\end{figure}

While recent interest in NRC has been motivated by a possibility of the \textit{intrinsic} origin of the effect, NRC naturally arises in multiply-connected superconductors. In superconducting loops, the critical current is modulated by an external flux $\phi=B S_{loop}$ piercing the loop. In a loop with asymmetric arms, the current maximum is shifted from $B=0$, and the sign of the shift depends on the direction of the current as shown in Fig.~\ref{f3}(a). A non-reciprocal component of the switching current \di\ is linear in $B$ in the vicinity of $B=0$, reaches extrema at $\phi\approx\phi_0/4$, changes sign and oscillates with a period $\Delta\phi=\phi_0$. Thus, an asymmetric loop is the simplest ``superconducting diode''.  There is a clear similarity between \di\ measured in an asymmetric superconducting loop and in an Al/InAs nanowire as emphasized in Fig.~\ref{f3}(b), suggesting that non-monotonic NRC in our nanowires may be due to emerging current loops.

\begin{figure}[t]
\centering
\includegraphics[width=.9\textwidth]{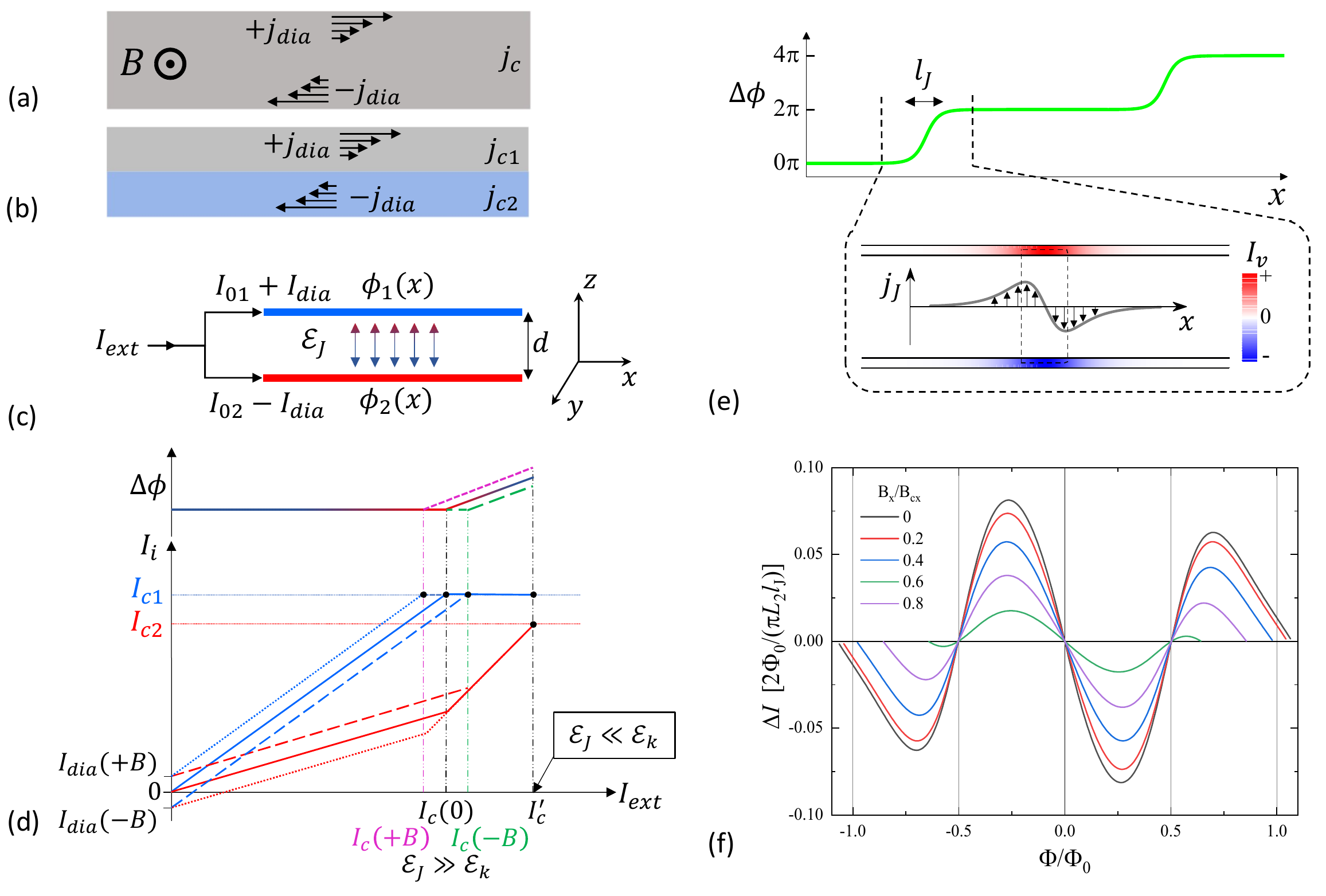}
\caption{\textbf{Non-reciprocity of the critical current in the presence of diamagnetic currents.} Diamagnetic currents in (a) a uniform superconductor and (b) a heterostructure. (c) A two-layer heterostructure is modeled as two zero thickness wires separated by a distance $d$ with coupling described by the Josephson energy $\mathcal{E}_J$. (d) Schematic of current distribution between the wires $I_i=I_{0i}-(-1)^i I_{dia}$, $i=1, 2$, and the phase difference $\Delta\phi$ as a function of an external current $I_{ext}=I_1+I_2$ for $B=0$ (solid lines), $B>0$ (dotted lines) and $B<0$ (dashed lines). For weakly coupled wires $\mathcal{E}_J\ll\mathcal{E}_k$, the critical current is field-independent $I^{'}_c=I_{c1}+I_{c2}$, see the text; the critical current is reduced and acquires a linear-in-$B$ correction in a strong coupling regime $\mathcal{E}_J\gg\mathcal{E}_k$ due to the phase locking $\Delta\phi=0$. (e) In the intermediate coupling regime $\mathcal{E}_J\sim\mathcal{E}_k$ Josephson vortices may form generating a $2\pi$ phase twist, in this case NRC becomes a non-monotonic function of $B$. (f) Calculated NRC $\Delta I$ is plotted as a function of flux $\Phi=S_v B_{y}$ for several $B_{x}$, Eq.~(\ref{eq:DeltaIFinal}).}
\label{fig:ring}
\label{f4}
\end{figure}

External magnetic field generates circular diamagnetic currents in a superconductor, as shown schematically in Fig.~\ref{f4}(a,b), and these currents affect the critical current. In homogeneous superconductors the presence of diamagnetic currents will not result in the critical current non-reciprocity, but in a heterogeneous superconductor, in general, their presence will lead to NRC. Qualitatively, the origin of NRC can be understood from a simplified model of a superconductor heterostructure represented as two coupled zero thickness superconducting wires separated by a distance $d$, Fig.~\ref{f4}(c). The total energy of the two-wire system can be written as a sum of kinetic and Josephson energies,
\begin{equation}
E_{\text{tot}}=\int dx \left[\mathcal{E}_{k}-\mathcal{E}_{J}\cos(\Delta\phi)\right]  \,,
\label{En_tot}
\end{equation}
where $\mathcal{E}_{k}=L_1I_1^2+L_2I_2^2$, $\mathcal{E}_{J}$ is the Josephson coupling, $\Delta\phi=\phi_{1}(x)-\phi_{2}(x)$ is the phase difference between superconducting condensates, and $L_i$ are the kinetic inductances per unit length in wires labeled by an index $i=1,2$. Supercurrents in each wire  $I_i=(2eL_i)^{-1}(\hbar\partial_x \phi_i-2eA_x)$ should satisfy  charge conservation constraint $I_1(x)+I_2(x)=I_{ext}$, where $I_{ext}$ is the applied external current. Detailed solution for this model can be found in the Supplementary Materials, and we outline now the main results. For small external currents ($I_1<I_{c1}$ and $I_2<I_{c2}$, where $I_{ci}$ are the critical currents in the wires) it is energetically favorable to keep the phase difference $\Delta\phi$ constant ($\Delta\phi=0$ for $\mathcal{E}_{J}>0$). Then, the currents can be expressed as $I_1=I_{01}+I_{dia}$ and $I_2=I_{02}-I_{dia}$, where $I_{01}, I_{02} \propto I_{ext}$ with $I_{01}/I_{02}=L_2/L_1=\eta^{-1}$ and $I_{dia}= B_y d/(L_1+L_2)$.  Dependence of $I_1$ and $I_2$ on $I_{ext}$ for $B_y>0$, $B_y=0$ and $B_y<0$ is plotted schematically in Fig.~\ref{f4}(d). As $I_{ext}$ increases and one of the currents ($I_1$ in our example) reaches the critical value $I_{c1}$, further external current increase requires an increase of $|\Delta\phi|$ because the excess current has to flow through the remaining superconducting wire with the current $I_2$.
In the case of weak interwire coupling, $\mathcal{E}_{J}\ll\mathcal{E}_{k}$, deviation of $\Delta\phi$ from zero does not lead to a significant energy penalty and the critical current of the whole system $I'_c=I_{c1}+I_{c2}$ does not depend on the magnetic field direction. In the opposite regime of strong coupling, $\mathcal{E}_{J}\gg\mathcal{E}_{k}$, the energy cost associated with the formation of Josephson currents (the last term in Eq.~\eqref{En_tot}) is prohibitively high and the whole system transitions to a normal state at $I_{ext}\approx(1+\eta)(I_{c1}-I_{dia})$, resulting in $\Delta I=-2(\eta+1)I_{dia}(B_y)$ (this equation is correct for $\beta>\eta + (\eta+1)I_{dia}/I_{c1}$, where $\beta = I_{c2}/I_{c1}$, NRC for other scenarios is listed in the Supplementary Materials). \textit{Thus a superconducting diode effect is a generic property of coupled multilayer superconductors.}

As $B_y$ and diamagnetic currents increase, the phase locking condition $\Delta\phi=0$ along the length of the wires leads to a significant increase of $\mathcal{E}_k$. At a critical field $B_c=(3/\pi^2)\Phi_0/( l_J  d)$  it becomes energetically favorable to reduce the overall energy by twisting the phase difference by $2\pi$ locally forming a Josephson vortex ($l_J\approx \Phi_0/(2\pi\sqrt{2\mathcal{E}_{J}L_2})$ and $\Phi_0=h/2e$ is the flux quantum). Evolution of the phase difference $\Delta\phi(x)=4 \arctan[\exp(x/l_J)]$, vortex-induced currents in the wires $I_v(x)$, and interwire Josephson current density $j_J(x)$ across a vortex are shown schematically in Fig.~\ref{f4}(e).
The maximum of $I_v(x)$ at the center of the  vortex determines the Josephson vortex contribution to NRC.
In the absence of quantum fluctuations formation of a vortex is accompanied by an abrupt re-distribution of currents between the wires, which results in a sawtooth NRC dependence on magnetic field. Generation of multiple Josephson vortices does not modify NRC compared to a single vortex case unless the vortices significantly overlap so that the maximum of $I_v(x)$ exceeds its single-vortex value.

In Fig.~\ref{f4}(f) we plot $\Delta I(B_y)$ for several $B_x$ using Eq.~\eqref{eq:DeltaIFinal} in the Supplemental Material. A gradual change of $\Delta I$ near $\Phi_0/2$ is due to quantum fluctuations of the winding number due to strong coupling of the vortex to current-carrying wires. This smearing is similar to the gradual change of a critical current in a ring connected to superconducting leads (Fig.~\ref{f3}), as compared to an abrupt reversal of persistent currents at $\Phi_0/2$ in isolated rings~\cite{Geim1997a}. The period of oscillations of $\Delta I$ corresponds to the flux threading an effective vortex area $S_v= (\pi^2/3) l_Jd=l_vd$. The period $\Delta B_\bot=400$ mT translates into the length $l_v\approx 500 \mathrm{nm}$, where $\Delta\phi$ substantially deviates from zero. We estimate  $l_v < \xi_{InAs}=\sqrt{\xi^0_{InAs} l^{m}_{InAs}}\approx 750$ nm and expect proximity-induced superconductivity in InAs to be preserved in the presence of a vortex. Here we use $\xi_{InAs}^0=\hbar v_F/\pi \Delta^*\approx 1.8\  \mu$m, induced gap in InAs $\Delta^*\approx\Delta = 1.796 k_BT_c = 230\mu eV$ (induced gap is close to the gap of Al in these heterostructures \cite{Nichele2017}), and the mean free path in uncapped InAs 2D gas $l^{m}_{InAs}\approx 300$ nm.

Finally, we use the two-wire model to estimate the temperature and in-plane field $B_\| \| \hat{x}$ dependences of NRC assuming that both parameters affect the Cooper pair density $n_2$ in InAs.  In the vicinity of $B_\bot=0$ the amplitude of $\Delta I\propto L_2^{-1}\propto n_2$ and is expected to decrease with an increase of $T$ or $B_\|$. The critical field $B_c\propto \sqrt{\mathcal{E}_{J}L_2}$ depends on $\mathcal{E}_{J}\propto n_2$, and the period of oscillations is expected to be $T$- and $B_\|$-independent, Fig.~\ref{f4}(d). Josephson coupling $\mathcal{E}_{J}$ is exponentially sensitive to the thickness of the InGaAs spacer and we expect slight variations of the period $\Delta B_\bot$ between the samples. These qualitative estimates are consistent with experimental observations, see Figs.~\ref{sf2} and ~\ref{sf3} in the Supplemental Material.

\section*{Methods}

\textbf{Materials.} The wafer was grown using Molecular Beam Epitaxy (MBE) on an InP substrate. The heterostructure consists of 1 $\mu$m graded In$_x$Al$_{1-x}$As insulating buffer followed by a In$_{0.75}$Ga$_{0.25}$As(4nm)/InAs(7nm)/In$_{0.75}$Ga$_{0.25}$As(10nm) multilayer structure capped \textit{in-situ} with 7nm of Al. The two-dimensional electron gas has a peak mobility of 28000 cm$^2$/Vs at a density $8 \times 10^{11}$ cm$^{-2}$.

\textbf{Sample Fabrication.}
The nanowires were fabricated using standard electron beam lithography. The mesas were defined by first removing the top Al layer with Al etchant Transene D and then a deep wet etching using H$_3$PO$_4$:H$_2$O$_2$:H$_2$O:C$_6$H$_8$O$_7$ (1ml:8ml:85ml:2g). Nanowires are defined in the second step of lithography by patterning the Al layer. Some devices have a top electrostatic gate, in these devices a 20 nm HfO$_2$ is grown by atomic layer deposition followed by a deposition of a Ti/Au (10/100 nm) gate.

\textbf{Measurements.}
Current-voltage sweeps were performed using a homemade high speed high resolution DAC/ADC (digital-to-analog and analog-to-digital converter) setup. The sweeps were automatically interrupted at the superconductor-normal transition ($I_{sw}$) in order to minimize device heating. Current sweep rate and delay between sweeps have been optimized to obtain $<5$ nA current resolution and to keep device temperature $<50$ mK at the base temperature of the fridge. The data has been corrected for an instrumental cooldown-dependent constant current offset (generated in the circuit by uncompensated voltages in the system and limited by a 100 k$\Omega$ current source resistor) to insure that $\Delta I=0$ at $B=0$.

\section*{Data Availability}

The data that support the findings of this study are available from the corresponding author upon reasonable request.

\section*{Acknowledgments}

The Al/InAs heterostructures were provided by Michael Manfra Group at Purdue University. The authors thank Lev Ioffe and Igor Aleiner for stimulating discussions. Experimental part was initially supported by the U.S. Department of Energy, Office of Basic Energy Sciences, Division of Materials Sciences and Engineering under Award DE-SC0008630, the work was completed with the support by NSF award DMR-DMR-2005092 (A.S. and L.P.R.).  Theoretical work is supported by the U.S. Department of Energy, Office of Basic Energy Sciences, Division of Materials Sciences and Engineering under Award DE-SC0010544 (Y.L-G) and the Office of the Under Secretary of Defense for Research and Engineering under award number FA9550-22-1-0354 (J.I.V.).

\section*{Author contributions}

L.P.R conceived, A.S. performed experiments, J.I.V. and Y.L.G developed the theory. All authors participated in writing the manuscript.


\begin{thebibliography}{40}%
\makeatletter
\providecommand \@ifxundefined [1]{%
 \@ifx{#1\undefined}
}%
\providecommand \@ifnum [1]{%
 \ifnum #1\expandafter \@firstoftwo
 \else \expandafter \@secondoftwo
 \fi
}%
\providecommand \@ifx [1]{%
 \ifx #1\expandafter \@firstoftwo
 \else \expandafter \@secondoftwo
 \fi
}%
\providecommand \natexlab [1]{#1}%
\providecommand \enquote  [1]{``#1''}%
\providecommand \bibnamefont  [1]{#1}%
\providecommand \bibfnamefont [1]{#1}%
\providecommand \citenamefont [1]{#1}%
\providecommand \href@noop [0]{\@secondoftwo}%
\providecommand \href [0]{\begingroup \@sanitize@url \@href}%
\providecommand \@href[1]{\@@startlink{#1}\@@href}%
\providecommand \@@href[1]{\endgroup#1\@@endlink}%
\providecommand \@sanitize@url [0]{\catcode `\\12\catcode `\$12\catcode
  `\&12\catcode `\#12\catcode `\^12\catcode `\_12\catcode `\%12\relax}%
\providecommand \@@startlink[1]{}%
\providecommand \@@endlink[0]{}%
\providecommand \url  [0]{\begingroup\@sanitize@url \@url }%
\providecommand \@url [1]{\endgroup\@href {#1}{\urlprefix }}%
\providecommand \urlprefix  [0]{URL }%
\providecommand \Eprint [0]{\href }%
\providecommand \doibase [0]{https://doi.org/}%
\providecommand \selectlanguage [0]{\@gobble}%
\providecommand \bibinfo  [0]{\@secondoftwo}%
\providecommand \bibfield  [0]{\@secondoftwo}%
\providecommand \translation [1]{[#1]}%
\providecommand \BibitemOpen [0]{}%
\providecommand \bibitemStop [0]{}%
\providecommand \bibitemNoStop [0]{.\EOS\space}%
\providecommand \EOS [0]{\spacefactor3000\relax}%
\providecommand \BibitemShut  [1]{\csname bibitem#1\endcsname}%
\let\auto@bib@innerbib\@empty
\bibitem [{\citenamefont {Ando}\ \emph {et~al.}(2020)\citenamefont {Ando},
  \citenamefont {Miyasaka}, \citenamefont {Li}, \citenamefont {Ishizuka},
  \citenamefont {Arakawa}, \citenamefont {Shiota}, \citenamefont {Moriyama},
  \citenamefont {Yanase},\ and\ \citenamefont {Ono}}]{Ando2020}%
  \BibitemOpen
  \bibfield  {author} {\bibinfo {author} {\bibfnamefont {F.}~\bibnamefont
  {Ando}}, \bibinfo {author} {\bibfnamefont {Y.}~\bibnamefont {Miyasaka}},
  \bibinfo {author} {\bibfnamefont {T.}~\bibnamefont {Li}}, \bibinfo {author}
  {\bibfnamefont {J.}~\bibnamefont {Ishizuka}}, \bibinfo {author}
  {\bibfnamefont {T.}~\bibnamefont {Arakawa}}, \bibinfo {author} {\bibfnamefont
  {Y.}~\bibnamefont {Shiota}}, \bibinfo {author} {\bibfnamefont
  {T.}~\bibnamefont {Moriyama}}, \bibinfo {author} {\bibfnamefont
  {Y.}~\bibnamefont {Yanase}},\ and\ \bibinfo {author} {\bibfnamefont
  {T.}~\bibnamefont {Ono}},\ }\bibfield  {title} {\bibinfo {title}
  {{Observation of superconducting diode effect}},\ }\href
  {https://doi.org/10.1038/s41586-020-2590-4} {\bibfield  {journal} {\bibinfo
  {journal} {Nature}\ }\textbf {\bibinfo {volume} {584}},\ \bibinfo {pages}
  {373} (\bibinfo {year} {2020})}\BibitemShut {NoStop}%
\bibitem [{\citenamefont {Agterberg}(2012)}]{Agterberg2012}%
  \BibitemOpen
  \bibfield  {author} {\bibinfo {author} {\bibfnamefont {D.~F.}\ \bibnamefont
  {Agterberg}},\ }\bibfield  {title} {\bibinfo {title} {Magnetoelectric
  effects, helical phases, and {FFLO} phases},\ }in\ \href
  {https://doi.org/10.1007/978-3-642-24624-1_5} {\emph {\bibinfo {booktitle}
  {Non-centrosymmetric Superconductors}}},\ \bibinfo {editor} {edited by\
  \bibinfo {editor} {\bibfnamefont {E.}~\bibnamefont {Bauer}}\ and\ \bibinfo
  {editor} {\bibfnamefont {M.}~\bibnamefont {Sigrist}}}\ (\bibinfo  {publisher}
  {Springer},\ \bibinfo {year} {2012})\ Chap.~\bibinfo {chapter} {5}, pp.\
  \bibinfo {pages} {155--170}\BibitemShut {NoStop}%
\bibitem [{\citenamefont {Daido}\ \emph {et~al.}(2022)\citenamefont {Daido},
  \citenamefont {Ikeda},\ and\ \citenamefont {Yanase}}]{Daido2022}%
  \BibitemOpen
  \bibfield  {author} {\bibinfo {author} {\bibfnamefont {A.}~\bibnamefont
  {Daido}}, \bibinfo {author} {\bibfnamefont {Y.}~\bibnamefont {Ikeda}},\ and\
  \bibinfo {author} {\bibfnamefont {Y.}~\bibnamefont {Yanase}},\ }\bibfield
  {title} {\bibinfo {title} {{Intrinsic Superconducting Diode Effect}},\ }\href
  {https://doi.org/10.1103/PhysRevLett.128.037001} {\bibfield  {journal}
  {\bibinfo  {journal} {Phys. Rev. Lett.}\ }\textbf {\bibinfo {volume} {128}},\
  \bibinfo {pages} {037001} (\bibinfo {year} {2022})}\BibitemShut {NoStop}%
\bibitem [{\citenamefont {He}\ \emph {et~al.}(2022)\citenamefont {He},
  \citenamefont {Tanaka},\ and\ \citenamefont {Nagaosa}}]{He2022}%
  \BibitemOpen
  \bibfield  {author} {\bibinfo {author} {\bibfnamefont {J.~J.}\ \bibnamefont
  {He}}, \bibinfo {author} {\bibfnamefont {Y.}~\bibnamefont {Tanaka}},\ and\
  \bibinfo {author} {\bibfnamefont {N.}~\bibnamefont {Nagaosa}},\ }\bibfield
  {title} {\bibinfo {title} {{A phenomenological theory of superconductor
  diodes}},\ }\href {https://doi.org/10.1088/1367-2630/ac6766} {\bibfield
  {journal} {\bibinfo  {journal} {New Journal of Physics}\ }\textbf {\bibinfo
  {volume} {24}},\ \bibinfo {pages} {053014} (\bibinfo {year}
  {2022})}\BibitemShut {NoStop}%
\bibitem [{\citenamefont {Burlakov}\ \emph {et~al.}(2014)\citenamefont
  {Burlakov}, \citenamefont {Gurtovoi}, \citenamefont {Il'in}, \citenamefont
  {Nikulov},\ and\ \citenamefont {Tulin}}]{Burlakov2014}%
  \BibitemOpen
  \bibfield  {author} {\bibinfo {author} {\bibfnamefont {A.~A.}\ \bibnamefont
  {Burlakov}}, \bibinfo {author} {\bibfnamefont {V.~L.}\ \bibnamefont
  {Gurtovoi}}, \bibinfo {author} {\bibfnamefont {A.~I.}\ \bibnamefont {Il'in}},
  \bibinfo {author} {\bibfnamefont {A.~V.}\ \bibnamefont {Nikulov}},\ and\
  \bibinfo {author} {\bibfnamefont {V.~A.}\ \bibnamefont {Tulin}},\ }\bibfield
  {title} {\bibinfo {title} {{Superconducting quantum interference device
  without Josephson junctions}},\ }\href
  {https://doi.org/10.1134/S0021364014030059} {\bibfield  {journal} {\bibinfo
  {journal} {JETP Lett.}\ }\textbf {\bibinfo {volume} {99}},\ \bibinfo {pages}
  {169} (\bibinfo {year} {2014})}\BibitemShut {NoStop}%
\bibitem [{\citenamefont {Rikken}\ and\ \citenamefont
  {Wyder}(2005)}]{Rikken2005}%
  \BibitemOpen
  \bibfield  {author} {\bibinfo {author} {\bibfnamefont {G.~L. J.~A.}\
  \bibnamefont {Rikken}}\ and\ \bibinfo {author} {\bibfnamefont
  {P.}~\bibnamefont {Wyder}},\ }\bibfield  {title} {\bibinfo {title}
  {{Magnetoelectric Anisotropy in Diffusive Transport}},\ }\href
  {https://doi.org/10.1103/PhysRevLett.94.016601} {\bibfield  {journal}
  {\bibinfo  {journal} {Phys. Rev. Lett.}\ }\textbf {\bibinfo {volume} {94}},\
  \bibinfo {pages} {016601} (\bibinfo {year} {2005})}\BibitemShut {NoStop}%
\bibitem [{\citenamefont {Levitov}\ \emph {et~al.}(1985)\citenamefont
  {Levitov}, \citenamefont {Nazarov},\ and\ \citenamefont
  {Eliashberg}}]{Levitov1985}%
  \BibitemOpen
  \bibfield  {author} {\bibinfo {author} {\bibfnamefont {L.~S.}\ \bibnamefont
  {Levitov}}, \bibinfo {author} {\bibfnamefont {Y.~V.}\ \bibnamefont
  {Nazarov}},\ and\ \bibinfo {author} {\bibfnamefont {G.~M.}\ \bibnamefont
  {Eliashberg}},\ }\bibfield  {title} {\bibinfo {title} {{Magnetostatics of
  Superconductors without an inversion center}},\ }\href@noop {} {\bibfield
  {journal} {\bibinfo  {journal} {JETP Lett.}\ }\textbf {\bibinfo {volume}
  {41}},\ \bibinfo {pages} {445} (\bibinfo {year} {1985})}\BibitemShut
  {NoStop}%
\bibitem [{\citenamefont {Edelstein}(1996)}]{Edelstein1996}%
  \BibitemOpen
  \bibfield  {author} {\bibinfo {author} {\bibfnamefont {V.~M.}\ \bibnamefont
  {Edelstein}},\ }\bibfield  {title} {\bibinfo {title} {{The Ginzburg - Landau
  equation for superconductors of polar symmetry}},\ }\href
  {https://doi.org/10.1088/0953-8984/8/3/012} {\bibfield  {journal} {\bibinfo
  {journal} {J. Phys. Condens. Matter}\ }\textbf {\bibinfo {volume} {8}},\
  \bibinfo {pages} {339} (\bibinfo {year} {1996})}\BibitemShut {NoStop}%
\bibitem [{\citenamefont {Ili\'{c}}\ and\ \citenamefont
  {Bergeret}(2022)}]{Ilic2022}%
  \BibitemOpen
  \bibfield  {author} {\bibinfo {author} {\bibfnamefont {S.}~\bibnamefont
  {Ili\'{c}}}\ and\ \bibinfo {author} {\bibfnamefont {F.~S.}\ \bibnamefont
  {Bergeret}},\ }\bibfield  {title} {\bibinfo {title} {Theory of the
  supercurrent diode effect in {R}ashba superconductors with arbitrary
  disorder},\ }\href {https://doi.org/10.1103/PhysRevLett.128.177001}
  {\bibfield  {journal} {\bibinfo  {journal} {Physical Review Letters}\
  }\textbf {\bibinfo {volume} {128}},\ \bibinfo {pages} {177001} (\bibinfo
  {year} {2022})}\BibitemShut {NoStop}%
\bibitem [{\citenamefont {Lyanda-Geller}\ \emph {et~al.}()\citenamefont
  {Lyanda-Geller}, \citenamefont {V\"ayrynen}, \citenamefont {Sundaresh},\ and\
  \citenamefont {Rokhinson}}]{Lyanda-Geller2022}%
  \BibitemOpen
  \bibfield  {author} {\bibinfo {author} {\bibfnamefont {Y.}~\bibnamefont
  {Lyanda-Geller}}, \bibinfo {author} {\bibfnamefont {J.~I.}\ \bibnamefont
  {V\"ayrynen}}, \bibinfo {author} {\bibfnamefont {A.}~\bibnamefont
  {Sundaresh}},\ and\ \bibinfo {author} {\bibfnamefont {L.~P.}\ \bibnamefont
  {Rokhinson}},\ }\href@noop {} {}\bibinfo {note} {To be published}\BibitemShut
  {NoStop}%
\bibitem [{\citenamefont {Vodolazov}\ \emph {et~al.}(2018)\citenamefont
  {Vodolazov}, \citenamefont {Aladyshkin}, \citenamefont {Pestov},
  \citenamefont {Vdovichev}, \citenamefont {Ustavshikov}, \citenamefont
  {Levichev}, \citenamefont {Putilov}, \citenamefont {Yunin}, \citenamefont
  {El'kina}, \citenamefont {Bukharov},\ and\ \citenamefont
  {Klushin}}]{Vodolazov2018}%
  \BibitemOpen
  \bibfield  {author} {\bibinfo {author} {\bibfnamefont {D.~Y.}\ \bibnamefont
  {Vodolazov}}, \bibinfo {author} {\bibfnamefont {A.~Y.}\ \bibnamefont
  {Aladyshkin}}, \bibinfo {author} {\bibfnamefont {E.~E.}\ \bibnamefont
  {Pestov}}, \bibinfo {author} {\bibfnamefont {S.~N.}\ \bibnamefont
  {Vdovichev}}, \bibinfo {author} {\bibfnamefont {S.~S.}\ \bibnamefont
  {Ustavshikov}}, \bibinfo {author} {\bibfnamefont {M.~Y.}\ \bibnamefont
  {Levichev}}, \bibinfo {author} {\bibfnamefont {A.~V.}\ \bibnamefont
  {Putilov}}, \bibinfo {author} {\bibfnamefont {P.~A.}\ \bibnamefont {Yunin}},
  \bibinfo {author} {\bibfnamefont {A.~I.}\ \bibnamefont {El'kina}}, \bibinfo
  {author} {\bibfnamefont {N.~N.}\ \bibnamefont {Bukharov}},\ and\ \bibinfo
  {author} {\bibfnamefont {A.~M.}\ \bibnamefont {Klushin}},\ }\bibfield
  {title} {\bibinfo {title} {{Peculiar superconducting properties of a thin
  film superconductor–normal metal bilayer with large ratio of
  resistivities}},\ }\href {https://doi.org/10.1088/1361-6668/aada2e}
  {\bibfield  {journal} {\bibinfo  {journal} {Supercond. Sci. Technol.}\
  }\textbf {\bibinfo {volume} {31}},\ \bibinfo {pages} {115004} (\bibinfo
  {year} {2018})}\BibitemShut {NoStop}%
\bibitem [{\citenamefont {Zinkl}\ \emph {et~al.}(2021)\citenamefont {Zinkl},
  \citenamefont {Hamamoto},\ and\ \citenamefont {Sigrist}}]{Zinkl2021}%
  \BibitemOpen
  \bibfield  {author} {\bibinfo {author} {\bibfnamefont {B.}~\bibnamefont
  {Zinkl}}, \bibinfo {author} {\bibfnamefont {K.}~\bibnamefont {Hamamoto}},\
  and\ \bibinfo {author} {\bibfnamefont {M.}~\bibnamefont {Sigrist}},\
  }\bibfield  {title} {\bibinfo {title} {{Symmetry conditions for the
  superconducting diode effect in chiral superconductors}},\ }\Eprint
  {https://arxiv.org/abs/2111.05340} {arXiv:2111.05340}  (\bibinfo {year}
  {2021})\BibitemShut {NoStop}%
\bibitem [{\citenamefont {Narita}\ \emph {et~al.}(2022)\citenamefont {Narita},
  \citenamefont {Ishizuka}, \citenamefont {Kawarazaki}, \citenamefont {Kan},
  \citenamefont {Shiota}, \citenamefont {Moriyama}, \citenamefont {Shimakawa},
  \citenamefont {Ognev}, \citenamefont {Samardak}, \citenamefont {Yanase},\
  and\ \citenamefont {Ono}}]{Narita2022}%
  \BibitemOpen
  \bibfield  {author} {\bibinfo {author} {\bibfnamefont {H.}~\bibnamefont
  {Narita}}, \bibinfo {author} {\bibfnamefont {J.}~\bibnamefont {Ishizuka}},
  \bibinfo {author} {\bibfnamefont {R.}~\bibnamefont {Kawarazaki}}, \bibinfo
  {author} {\bibfnamefont {D.}~\bibnamefont {Kan}}, \bibinfo {author}
  {\bibfnamefont {Y.}~\bibnamefont {Shiota}}, \bibinfo {author} {\bibfnamefont
  {T.}~\bibnamefont {Moriyama}}, \bibinfo {author} {\bibfnamefont
  {Y.}~\bibnamefont {Shimakawa}}, \bibinfo {author} {\bibfnamefont {A.~V.}\
  \bibnamefont {Ognev}}, \bibinfo {author} {\bibfnamefont {A.~S.}\ \bibnamefont
  {Samardak}}, \bibinfo {author} {\bibfnamefont {Y.}~\bibnamefont {Yanase}},\
  and\ \bibinfo {author} {\bibfnamefont {T.}~\bibnamefont {Ono}},\ }\bibfield
  {title} {\bibinfo {title} {Field-free superconducting diode effect in
  noncentrosymmetric superconductor/ferromagnet multilayers},\ }\href
  {https://doi.org/10.1038/s41565-022-01159-4} {\bibfield  {journal} {\bibinfo
  {journal} {Nature Nanotechnology}\ }\textbf {\bibinfo {volume} {17}},\
  \bibinfo {pages} {823} (\bibinfo {year} {2022})}\BibitemShut {NoStop}%
\bibitem [{\citenamefont {Bauriedl}\ \emph {et~al.}(2022)\citenamefont
  {Bauriedl}, \citenamefont {B{\"{a}}uml}, \citenamefont {Fuchs}, \citenamefont
  {Baumgartner}, \citenamefont {Paulik}, \citenamefont {Bauer}, \citenamefont
  {Lin}, \citenamefont {Lupton}, \citenamefont {Taniguchi}, \citenamefont
  {Watanabe}, \citenamefont {Strunk},\ and\ \citenamefont
  {Paradiso}}]{Bauriedl2022}%
  \BibitemOpen
  \bibfield  {author} {\bibinfo {author} {\bibfnamefont {L.}~\bibnamefont
  {Bauriedl}}, \bibinfo {author} {\bibfnamefont {C.}~\bibnamefont
  {B{\"{a}}uml}}, \bibinfo {author} {\bibfnamefont {L.}~\bibnamefont {Fuchs}},
  \bibinfo {author} {\bibfnamefont {C.}~\bibnamefont {Baumgartner}}, \bibinfo
  {author} {\bibfnamefont {N.}~\bibnamefont {Paulik}}, \bibinfo {author}
  {\bibfnamefont {J.~M.}\ \bibnamefont {Bauer}}, \bibinfo {author}
  {\bibfnamefont {K.-q.}\ \bibnamefont {Lin}}, \bibinfo {author} {\bibfnamefont
  {J.~M.}\ \bibnamefont {Lupton}}, \bibinfo {author} {\bibfnamefont
  {T.}~\bibnamefont {Taniguchi}}, \bibinfo {author} {\bibfnamefont
  {K.}~\bibnamefont {Watanabe}}, \bibinfo {author} {\bibfnamefont
  {C.}~\bibnamefont {Strunk}},\ and\ \bibinfo {author} {\bibfnamefont
  {N.}~\bibnamefont {Paradiso}},\ }\bibfield  {title} {\bibinfo {title}
  {{Supercurrent diode effect and magnetochiral anisotropy in few-layer
  NbSe2}},\ }\bibfield  {journal} {\bibinfo  {journal} {Nature Communications}\
  }\href {https://doi.org/10.1038/s41467-022-31954-5}
  {10.1038/s41467-022-31954-5} (\bibinfo {year} {2022})\BibitemShut {NoStop}%
\bibitem [{\citenamefont {Lin}\ \emph {et~al.}(2022)\citenamefont {Lin},
  \citenamefont {Siriviboon}, \citenamefont {Scammell}, \citenamefont {Liu},
  \citenamefont {Rhodes}, \citenamefont {Watanabe}, \citenamefont {Taniguchi},
  \citenamefont {Hone}, \citenamefont {Scheurer},\ and\ \citenamefont
  {Li}}]{Jiang2022}%
  \BibitemOpen
  \bibfield  {author} {\bibinfo {author} {\bibfnamefont {J.-x.}\ \bibnamefont
  {Lin}}, \bibinfo {author} {\bibfnamefont {P.}~\bibnamefont {Siriviboon}},
  \bibinfo {author} {\bibfnamefont {H.~D.}\ \bibnamefont {Scammell}}, \bibinfo
  {author} {\bibfnamefont {S.}~\bibnamefont {Liu}}, \bibinfo {author}
  {\bibfnamefont {D.}~\bibnamefont {Rhodes}}, \bibinfo {author} {\bibfnamefont
  {K.}~\bibnamefont {Watanabe}}, \bibinfo {author} {\bibfnamefont
  {T.}~\bibnamefont {Taniguchi}}, \bibinfo {author} {\bibfnamefont
  {J.}~\bibnamefont {Hone}}, \bibinfo {author} {\bibfnamefont {M.~S.}\
  \bibnamefont {Scheurer}},\ and\ \bibinfo {author} {\bibfnamefont {J.~I.~A.}\
  \bibnamefont {Li}},\ }\bibfield  {title} {\bibinfo {title} {Zero-field
  superconducting diode effect in small-twist-angle trilayer graphene},\
  }\bibfield  {journal} {\bibinfo  {journal} {Nature Physics}\ }\href
  {https://doi.org/10.1038/s41567-022-01700-1} {10.1038/s41567-022-01700-1}
  (\bibinfo {year} {2022})\BibitemShut {NoStop}%
\bibitem [{\citenamefont {Lyu}\ \emph {et~al.}(2021)\citenamefont {Lyu},
  \citenamefont {Jiang}, \citenamefont {Wang}, \citenamefont {Xiao},
  \citenamefont {Dong}, \citenamefont {Chen}, \citenamefont
  {Milo{\v{s}}evi{\'{c}}}, \citenamefont {Wang}, \citenamefont {Divan},
  \citenamefont {Pearson}, \citenamefont {Wu}, \citenamefont {Peeters},\ and\
  \citenamefont {Kwok}}]{Lyu2021}%
  \BibitemOpen
  \bibfield  {author} {\bibinfo {author} {\bibfnamefont {Y.-Y.}\ \bibnamefont
  {Lyu}}, \bibinfo {author} {\bibfnamefont {J.}~\bibnamefont {Jiang}}, \bibinfo
  {author} {\bibfnamefont {Y.-L.}\ \bibnamefont {Wang}}, \bibinfo {author}
  {\bibfnamefont {Z.-L.}\ \bibnamefont {Xiao}}, \bibinfo {author}
  {\bibfnamefont {S.}~\bibnamefont {Dong}}, \bibinfo {author} {\bibfnamefont
  {Q.-H.}\ \bibnamefont {Chen}}, \bibinfo {author} {\bibfnamefont {M.~V.}\
  \bibnamefont {Milo{\v{s}}evi{\'{c}}}}, \bibinfo {author} {\bibfnamefont
  {H.}~\bibnamefont {Wang}}, \bibinfo {author} {\bibfnamefont {R.}~\bibnamefont
  {Divan}}, \bibinfo {author} {\bibfnamefont {J.~E.}\ \bibnamefont {Pearson}},
  \bibinfo {author} {\bibfnamefont {P.}~\bibnamefont {Wu}}, \bibinfo {author}
  {\bibfnamefont {F.~M.}\ \bibnamefont {Peeters}},\ and\ \bibinfo {author}
  {\bibfnamefont {W.-K.}\ \bibnamefont {Kwok}},\ }\bibfield  {title} {\bibinfo
  {title} {{Superconducting diode effect via conformal-mapped nanoholes}},\
  }\href {https://doi.org/10.1038/s41467-021-23077-0} {\bibfield  {journal}
  {\bibinfo  {journal} {Nat. Commun.}\ }\textbf {\bibinfo {volume} {12}},\
  \bibinfo {pages} {2703} (\bibinfo {year} {2021})}\BibitemShut {NoStop}%
\bibitem [{\citenamefont {Suri}\ \emph {et~al.}(2022)\citenamefont {Suri},
  \citenamefont {Kamra}, \citenamefont {Meier}, \citenamefont {Kronseder},
  \citenamefont {Belzig}, \citenamefont {Back},\ and\ \citenamefont
  {Strunk}}]{Suri2022}%
  \BibitemOpen
  \bibfield  {author} {\bibinfo {author} {\bibfnamefont {D.}~\bibnamefont
  {Suri}}, \bibinfo {author} {\bibfnamefont {A.}~\bibnamefont {Kamra}},
  \bibinfo {author} {\bibfnamefont {T.~N.}\ \bibnamefont {Meier}}, \bibinfo
  {author} {\bibfnamefont {M.}~\bibnamefont {Kronseder}}, \bibinfo {author}
  {\bibfnamefont {W.}~\bibnamefont {Belzig}}, \bibinfo {author} {\bibfnamefont
  {C.~H.}\ \bibnamefont {Back}},\ and\ \bibinfo {author} {\bibfnamefont
  {C.}~\bibnamefont {Strunk}},\ }\bibfield  {title} {\bibinfo {title}
  {{Non-reciprocity of vortex-limited critical current in conventional
  superconducting micro-bridges}},\ }\bibfield  {journal} {\bibinfo  {journal}
  {Applied Physics Letters}\ }\textbf {\bibinfo {volume} {121}},\ \href
  {https://doi.org/10.1063/5.0109753} {10.1063/5.0109753} (\bibinfo {year}
  {2022})\BibitemShut {NoStop}%
\bibitem [{\citenamefont {Ustavschikov}\ \emph {et~al.}(2022)\citenamefont
  {Ustavschikov}, \citenamefont {Levichev}, \citenamefont {Pashenkin},
  \citenamefont {Gusev}, \citenamefont {Gusev},\ and\ \citenamefont
  {Vodolazov}}]{Ustavschikov2022}%
  \BibitemOpen
  \bibfield  {author} {\bibinfo {author} {\bibfnamefont {S.~S.}\ \bibnamefont
  {Ustavschikov}}, \bibinfo {author} {\bibfnamefont {M.~Y.}\ \bibnamefont
  {Levichev}}, \bibinfo {author} {\bibfnamefont {I.~Y.}\ \bibnamefont
  {Pashenkin}}, \bibinfo {author} {\bibfnamefont {N.~S.}\ \bibnamefont
  {Gusev}}, \bibinfo {author} {\bibfnamefont {S.~A.}\ \bibnamefont {Gusev}},\
  and\ \bibinfo {author} {\bibfnamefont {D.~Y.}\ \bibnamefont {Vodolazov}},\
  }\bibfield  {title} {\bibinfo {title} {{Diode Effect in a Superconducting
  Hybrid Cu/MoN Strip with a Lateral Cut}},\ }\href
  {https://doi.org/10.1134/S1063776122080064} {\bibfield  {journal} {\bibinfo
  {journal} {Journal of Experimental and Theoretical Physics}\ }\textbf
  {\bibinfo {volume} {135}},\ \bibinfo {pages} {226} (\bibinfo {year}
  {2022})}\BibitemShut {NoStop}%
\bibitem [{\citenamefont {Hou}\ \emph {et~al.}(2022)\citenamefont {Hou},
  \citenamefont {Nichele}, \citenamefont {Chi}, \citenamefont {Lodesani},
  \citenamefont {Wu}, \citenamefont {Ritter}, \citenamefont {Haxell},
  \citenamefont {Davydova}, \citenamefont {Ili{\'{c}}}, \citenamefont
  {Bergeret}, \citenamefont {Kamra}, \citenamefont {Fu}, \citenamefont {Lee},\
  and\ \citenamefont {Moodera}}]{Hou2022}%
  \BibitemOpen
  \bibfield  {author} {\bibinfo {author} {\bibfnamefont {Y.}~\bibnamefont
  {Hou}}, \bibinfo {author} {\bibfnamefont {F.}~\bibnamefont {Nichele}},
  \bibinfo {author} {\bibfnamefont {H.}~\bibnamefont {Chi}}, \bibinfo {author}
  {\bibfnamefont {A.}~\bibnamefont {Lodesani}}, \bibinfo {author}
  {\bibfnamefont {Y.}~\bibnamefont {Wu}}, \bibinfo {author} {\bibfnamefont
  {M.~F.}\ \bibnamefont {Ritter}}, \bibinfo {author} {\bibfnamefont {D.~Z.}\
  \bibnamefont {Haxell}}, \bibinfo {author} {\bibfnamefont {M.}~\bibnamefont
  {Davydova}}, \bibinfo {author} {\bibfnamefont {S.}~\bibnamefont
  {Ili{\'{c}}}}, \bibinfo {author} {\bibfnamefont {F.~S.}\ \bibnamefont
  {Bergeret}}, \bibinfo {author} {\bibfnamefont {A.}~\bibnamefont {Kamra}},
  \bibinfo {author} {\bibfnamefont {L.}~\bibnamefont {Fu}}, \bibinfo {author}
  {\bibfnamefont {P.~A.}\ \bibnamefont {Lee}},\ and\ \bibinfo {author}
  {\bibfnamefont {J.~S.}\ \bibnamefont {Moodera}},\ }\bibfield  {title}
  {\bibinfo {title} {{Ubiquitous Superconducting Diode Effect in Superconductor
  Thin Films}},\ }\Eprint {https://arxiv.org/abs/2205.09276} {arXiv:2205.09276}
   (\bibinfo {year} {2022})\BibitemShut {NoStop}%
\bibitem [{\citenamefont {Diez-Merida}\ \emph {et~al.}(2021)\citenamefont
  {Diez-Merida}, \citenamefont {Diez-Carlon}, \citenamefont {Yang},
  \citenamefont {Xie}, \citenamefont {Gao}, \citenamefont {Watanabe},
  \citenamefont {Taniguchi}, \citenamefont {Lu}, \citenamefont {Law},\ and\
  \citenamefont {Efetov}}]{Diez-Merida2021}%
  \BibitemOpen
  \bibfield  {author} {\bibinfo {author} {\bibfnamefont {J.}~\bibnamefont
  {Diez-Merida}}, \bibinfo {author} {\bibfnamefont {A.}~\bibnamefont
  {Diez-Carlon}}, \bibinfo {author} {\bibfnamefont {S.~Y.}\ \bibnamefont
  {Yang}}, \bibinfo {author} {\bibfnamefont {Y.~M.}\ \bibnamefont {Xie}},
  \bibinfo {author} {\bibfnamefont {X.~J.}\ \bibnamefont {Gao}}, \bibinfo
  {author} {\bibfnamefont {K.}~\bibnamefont {Watanabe}}, \bibinfo {author}
  {\bibfnamefont {T.}~\bibnamefont {Taniguchi}}, \bibinfo {author}
  {\bibfnamefont {X.}~\bibnamefont {Lu}}, \bibinfo {author} {\bibfnamefont
  {K.~T.}\ \bibnamefont {Law}},\ and\ \bibinfo {author} {\bibfnamefont {D.~K.}\
  \bibnamefont {Efetov}},\ }\bibfield  {title} {\bibinfo {title} {{Magnetic
  Josephson Junctions and Superconducting Diodes in Magic Angle Twisted Bilayer
  Graphene}},\ }\Eprint {https://arxiv.org/abs/2110.01067} {arXiv:2110.01067}
  (\bibinfo {year} {2021})\BibitemShut {NoStop}%
\bibitem [{\citenamefont {Baumgartner}\ \emph
  {et~al.}(2022{\natexlab{a}})\citenamefont {Baumgartner}, \citenamefont
  {Fuchs}, \citenamefont {Costa}, \citenamefont {Reinhardt}, \citenamefont
  {Gronin}, \citenamefont {Gardner}, \citenamefont {Lindemann}, \citenamefont
  {Manfra}, \citenamefont {{Faria Junior}}, \citenamefont {Kochan},
  \citenamefont {Fabian}, \citenamefont {Paradiso},\ and\ \citenamefont
  {Strunk}}]{Baumgartner2022}%
  \BibitemOpen
  \bibfield  {author} {\bibinfo {author} {\bibfnamefont {C.}~\bibnamefont
  {Baumgartner}}, \bibinfo {author} {\bibfnamefont {L.}~\bibnamefont {Fuchs}},
  \bibinfo {author} {\bibfnamefont {A.}~\bibnamefont {Costa}}, \bibinfo
  {author} {\bibfnamefont {S.}~\bibnamefont {Reinhardt}}, \bibinfo {author}
  {\bibfnamefont {S.}~\bibnamefont {Gronin}}, \bibinfo {author} {\bibfnamefont
  {G.~C.}\ \bibnamefont {Gardner}}, \bibinfo {author} {\bibfnamefont
  {T.}~\bibnamefont {Lindemann}}, \bibinfo {author} {\bibfnamefont {M.~J.}\
  \bibnamefont {Manfra}}, \bibinfo {author} {\bibfnamefont {P.~E.}\
  \bibnamefont {{Faria Junior}}}, \bibinfo {author} {\bibfnamefont
  {D.}~\bibnamefont {Kochan}}, \bibinfo {author} {\bibfnamefont
  {J.}~\bibnamefont {Fabian}}, \bibinfo {author} {\bibfnamefont
  {N.}~\bibnamefont {Paradiso}},\ and\ \bibinfo {author} {\bibfnamefont
  {C.}~\bibnamefont {Strunk}},\ }\bibfield  {title} {\bibinfo {title}
  {{Supercurrent rectification and magnetochiral effects in symmetric Josephson
  junctions}},\ }\href {https://doi.org/10.1038/s41565-021-01009-9} {\bibfield
  {journal} {\bibinfo  {journal} {Nature Nanotechnology}\ }\textbf {\bibinfo
  {volume} {17}},\ \bibinfo {pages} {39} (\bibinfo {year}
  {2022}{\natexlab{a}})}\BibitemShut {NoStop}%
\bibitem [{\citenamefont {Pal}\ \emph {et~al.}(2022)\citenamefont {Pal},
  \citenamefont {Chakraborty}, \citenamefont {Sivakumar}, \citenamefont
  {Davydova}, \citenamefont {Gopi}, \citenamefont {Pandeya}, \citenamefont
  {Krieger}, \citenamefont {Zhang}, \citenamefont {Date}, \citenamefont {Ju},
  \citenamefont {Yuan}, \citenamefont {Schr{\"{o}}ter}, \citenamefont {Fu},\
  and\ \citenamefont {Parkin}}]{Pal2022}%
  \BibitemOpen
  \bibfield  {author} {\bibinfo {author} {\bibfnamefont {B.}~\bibnamefont
  {Pal}}, \bibinfo {author} {\bibfnamefont {A.}~\bibnamefont {Chakraborty}},
  \bibinfo {author} {\bibfnamefont {P.~K.}\ \bibnamefont {Sivakumar}}, \bibinfo
  {author} {\bibfnamefont {M.}~\bibnamefont {Davydova}}, \bibinfo {author}
  {\bibfnamefont {A.~K.}\ \bibnamefont {Gopi}}, \bibinfo {author}
  {\bibfnamefont {A.~K.}\ \bibnamefont {Pandeya}}, \bibinfo {author}
  {\bibfnamefont {J.~A.}\ \bibnamefont {Krieger}}, \bibinfo {author}
  {\bibfnamefont {Y.}~\bibnamefont {Zhang}}, \bibinfo {author} {\bibfnamefont
  {M.}~\bibnamefont {Date}}, \bibinfo {author} {\bibfnamefont {S.}~\bibnamefont
  {Ju}}, \bibinfo {author} {\bibfnamefont {N.}~\bibnamefont {Yuan}}, \bibinfo
  {author} {\bibfnamefont {N.~B.~M.}\ \bibnamefont {Schr{\"{o}}ter}}, \bibinfo
  {author} {\bibfnamefont {L.}~\bibnamefont {Fu}},\ and\ \bibinfo {author}
  {\bibfnamefont {S.~S.~P.}\ \bibnamefont {Parkin}},\ }\bibfield  {title}
  {\bibinfo {title} {Josephson diode effect from cooper pair momentum in a
  topological semimetal},\ }\bibfield  {journal} {\bibinfo  {journal} {Nature
  Physics}\ }\href {https://doi.org/10.1038/s41567-022-01699-5}
  {10.1038/s41567-022-01699-5} (\bibinfo {year} {2022})\BibitemShut {NoStop}%
\bibitem [{\citenamefont {Jeon}\ \emph {et~al.}(2022)\citenamefont {Jeon},
  \citenamefont {Kim}, \citenamefont {Yoon}, \citenamefont {Jeon},
  \citenamefont {Han}, \citenamefont {Cottet}, \citenamefont {Kontos},\ and\
  \citenamefont {Parkin}}]{Jeon2022}%
  \BibitemOpen
  \bibfield  {author} {\bibinfo {author} {\bibfnamefont {K.~R.}\ \bibnamefont
  {Jeon}}, \bibinfo {author} {\bibfnamefont {J.~K.}\ \bibnamefont {Kim}},
  \bibinfo {author} {\bibfnamefont {J.}~\bibnamefont {Yoon}}, \bibinfo {author}
  {\bibfnamefont {J.~C.}\ \bibnamefont {Jeon}}, \bibinfo {author}
  {\bibfnamefont {H.}~\bibnamefont {Han}}, \bibinfo {author} {\bibfnamefont
  {A.}~\bibnamefont {Cottet}}, \bibinfo {author} {\bibfnamefont
  {T.}~\bibnamefont {Kontos}},\ and\ \bibinfo {author} {\bibfnamefont {S.~S.}\
  \bibnamefont {Parkin}},\ }\bibfield  {title} {\bibinfo {title} {{Zero-field
  polarity-reversible Josephson supercurrent diodes enabled by a
  proximity-magnetized Pt barrier}},\ }\bibfield  {journal} {\bibinfo
  {journal} {Nature Materials}\ }\textbf {\bibinfo {volume} {21}},\ \href
  {https://doi.org/10.1038/s41563-022-01300-7} {10.1038/s41563-022-01300-7}
  (\bibinfo {year} {2022})\BibitemShut {NoStop}%
\bibitem [{\citenamefont {Gupta}\ \emph {et~al.}(2022)\citenamefont {Gupta},
  \citenamefont {Graziano}, \citenamefont {Pendharkar}, \citenamefont {Dong},
  \citenamefont {Dempsey}, \citenamefont {Palmstr{\o}m},\ and\ \citenamefont
  {Pribiag}}]{Gupta2022}%
  \BibitemOpen
  \bibfield  {author} {\bibinfo {author} {\bibfnamefont {M.}~\bibnamefont
  {Gupta}}, \bibinfo {author} {\bibfnamefont {G.~V.}\ \bibnamefont {Graziano}},
  \bibinfo {author} {\bibfnamefont {M.}~\bibnamefont {Pendharkar}}, \bibinfo
  {author} {\bibfnamefont {J.~T.}\ \bibnamefont {Dong}}, \bibinfo {author}
  {\bibfnamefont {C.~P.}\ \bibnamefont {Dempsey}}, \bibinfo {author}
  {\bibfnamefont {C.}~\bibnamefont {Palmstr{\o}m}},\ and\ \bibinfo {author}
  {\bibfnamefont {V.~S.}\ \bibnamefont {Pribiag}},\ }\bibfield  {title}
  {\bibinfo {title} {{Superconducting Diode Effect in a Three-terminal
  Josephson Device}}} (\bibinfo {year} {2022})\BibitemShut {NoStop}%
\bibitem [{\citenamefont {Golod}\ and\ \citenamefont
  {Krasnov}(2022)}]{Golod2022}%
  \BibitemOpen
  \bibfield  {author} {\bibinfo {author} {\bibfnamefont {T.}~\bibnamefont
  {Golod}}\ and\ \bibinfo {author} {\bibfnamefont {V.~M.}\ \bibnamefont
  {Krasnov}},\ }\bibfield  {title} {\bibinfo {title} {{Demonstration of a
  superconducting diode-with-memory, operational at zero magnetic field with
  switchable nonreciprocity}},\ }\href
  {https://doi.org/10.1038/s41467-022-31256-w} {\bibfield  {journal} {\bibinfo
  {journal} {Nature Communications}\ }\textbf {\bibinfo {volume} {13}},\
  \bibinfo {pages} {1} (\bibinfo {year} {2022})}\BibitemShut {NoStop}%
\bibitem [{\citenamefont {Wu}\ \emph {et~al.}(2022)\citenamefont {Wu},
  \citenamefont {Wang}, \citenamefont {Xu}, \citenamefont {Sivakumar},
  \citenamefont {Pasco}, \citenamefont {Filippozzi}, \citenamefont {Parkin},
  \citenamefont {Zeng}, \citenamefont {McQueen},\ and\ \citenamefont
  {Ali}}]{Wu2022}%
  \BibitemOpen
  \bibfield  {author} {\bibinfo {author} {\bibfnamefont {H.}~\bibnamefont
  {Wu}}, \bibinfo {author} {\bibfnamefont {Y.}~\bibnamefont {Wang}}, \bibinfo
  {author} {\bibfnamefont {Y.}~\bibnamefont {Xu}}, \bibinfo {author}
  {\bibfnamefont {P.~K.}\ \bibnamefont {Sivakumar}}, \bibinfo {author}
  {\bibfnamefont {C.}~\bibnamefont {Pasco}}, \bibinfo {author} {\bibfnamefont
  {U.}~\bibnamefont {Filippozzi}}, \bibinfo {author} {\bibfnamefont {S.~S.}\
  \bibnamefont {Parkin}}, \bibinfo {author} {\bibfnamefont {Y.~J.}\
  \bibnamefont {Zeng}}, \bibinfo {author} {\bibfnamefont {T.}~\bibnamefont
  {McQueen}},\ and\ \bibinfo {author} {\bibfnamefont {M.~N.}\ \bibnamefont
  {Ali}},\ }\bibfield  {title} {\bibinfo {title} {{The field-free Josephson
  diode in a van der Waals heterostructure}},\ }\href
  {https://doi.org/10.1038/s41586-022-04504-8} {\bibfield  {journal} {\bibinfo
  {journal} {Nature}\ }\textbf {\bibinfo {volume} {604}},\ \bibinfo {pages}
  {653} (\bibinfo {year} {2022})}\BibitemShut {NoStop}%
\bibitem [{\citenamefont {Chiles}\ \emph {et~al.}(2022)\citenamefont {Chiles},
  \citenamefont {Arnault}, \citenamefont {Chen}, \citenamefont {Larson},
  \citenamefont {Zhao}, \citenamefont {Watanabe}, \citenamefont {Taniguchi},
  \citenamefont {Amet},\ and\ \citenamefont {Finkelstein}}]{John2022}%
  \BibitemOpen
  \bibfield  {author} {\bibinfo {author} {\bibfnamefont {J.}~\bibnamefont
  {Chiles}}, \bibinfo {author} {\bibfnamefont {E.~G.}\ \bibnamefont {Arnault}},
  \bibinfo {author} {\bibfnamefont {C.-C.}\ \bibnamefont {Chen}}, \bibinfo
  {author} {\bibfnamefont {T.~F.~Q.}\ \bibnamefont {Larson}}, \bibinfo {author}
  {\bibfnamefont {L.}~\bibnamefont {Zhao}}, \bibinfo {author} {\bibfnamefont
  {K.}~\bibnamefont {Watanabe}}, \bibinfo {author} {\bibfnamefont
  {T.}~\bibnamefont {Taniguchi}}, \bibinfo {author} {\bibfnamefont
  {F.}~\bibnamefont {Amet}},\ and\ \bibinfo {author} {\bibfnamefont
  {G.}~\bibnamefont {Finkelstein}},\ }\bibfield  {title} {\bibinfo {title}
  {{Non-Reciprocal Supercurrents in a Field-Free Graphene Josephson Triode}},\
  }\Eprint {https://arxiv.org/abs/2210.02644} {arXiv:2210.02644}  (\bibinfo
  {year} {2022})\BibitemShut {NoStop}%
\bibitem [{\citenamefont {Turini}\ \emph {et~al.}(2022)\citenamefont {Turini},
  \citenamefont {Salimian}, \citenamefont {Carrega}, \citenamefont {Iorio},
  \citenamefont {Strambini}, \citenamefont {Giazotto}, \citenamefont {Zannier},
  \citenamefont {Sorba},\ and\ \citenamefont {Heun}}]{Bianca2022}%
  \BibitemOpen
  \bibfield  {author} {\bibinfo {author} {\bibfnamefont {B.}~\bibnamefont
  {Turini}}, \bibinfo {author} {\bibfnamefont {S.}~\bibnamefont {Salimian}},
  \bibinfo {author} {\bibfnamefont {M.}~\bibnamefont {Carrega}}, \bibinfo
  {author} {\bibfnamefont {A.}~\bibnamefont {Iorio}}, \bibinfo {author}
  {\bibfnamefont {E.}~\bibnamefont {Strambini}}, \bibinfo {author}
  {\bibfnamefont {F.}~\bibnamefont {Giazotto}}, \bibinfo {author}
  {\bibfnamefont {V.}~\bibnamefont {Zannier}}, \bibinfo {author} {\bibfnamefont
  {L.}~\bibnamefont {Sorba}},\ and\ \bibinfo {author} {\bibfnamefont
  {S.}~\bibnamefont {Heun}},\ }\bibfield  {title} {\bibinfo {title} {{Josephson
  Diode Effect in High Mobility InSb Nanoflags}},\ }\bibfield  {journal}
  {\bibinfo  {journal} {Nano Leters}\ }\href
  {https://doi.org/10.1021/acs.nanolett.2c02899} {10.1021/acs.nanolett.2c02899}
  (\bibinfo {year} {2022})\BibitemShut {NoStop}%
\bibitem [{\citenamefont {Bardeen}(1962)}]{Bardeen1962}%
  \BibitemOpen
  \bibfield  {author} {\bibinfo {author} {\bibfnamefont {J.}~\bibnamefont
  {Bardeen}},\ }\bibfield  {title} {\bibinfo {title} {{Critical Fields and
  Currents in Superconductors}},\ }\href
  {https://doi.org/10.1103/RevModPhys.34.667} {\bibfield  {journal} {\bibinfo
  {journal} {Rev. Mod. Phys.}\ }\textbf {\bibinfo {volume} {34}},\ \bibinfo
  {pages} {667} (\bibinfo {year} {1962})}\BibitemShut {NoStop}%
\bibitem [{\citenamefont {Shabani}\ \emph {et~al.}(2016)\citenamefont
  {Shabani}, \citenamefont {Kjaergaard}, \citenamefont {Suominen},
  \citenamefont {Kim}, \citenamefont {Nichele}, \citenamefont {Pakrouski},
  \citenamefont {Stankevic}, \citenamefont {Lutchyn}, \citenamefont
  {Krogstrup}, \citenamefont {Feidenhans'l}, \citenamefont {Kraemer},
  \citenamefont {Nayak}, \citenamefont {Troyer}, \citenamefont {Marcus},\ and\
  \citenamefont {Palmstr\"om}}]{Shabani2016}%
  \BibitemOpen
  \bibfield  {author} {\bibinfo {author} {\bibfnamefont {J.}~\bibnamefont
  {Shabani}}, \bibinfo {author} {\bibfnamefont {M.}~\bibnamefont {Kjaergaard}},
  \bibinfo {author} {\bibfnamefont {H.~J.}\ \bibnamefont {Suominen}}, \bibinfo
  {author} {\bibfnamefont {Y.}~\bibnamefont {Kim}}, \bibinfo {author}
  {\bibfnamefont {F.}~\bibnamefont {Nichele}}, \bibinfo {author} {\bibfnamefont
  {K.}~\bibnamefont {Pakrouski}}, \bibinfo {author} {\bibfnamefont
  {T.}~\bibnamefont {Stankevic}}, \bibinfo {author} {\bibfnamefont {R.~M.}\
  \bibnamefont {Lutchyn}}, \bibinfo {author} {\bibfnamefont {P.}~\bibnamefont
  {Krogstrup}}, \bibinfo {author} {\bibfnamefont {R.}~\bibnamefont
  {Feidenhans'l}}, \bibinfo {author} {\bibfnamefont {S.}~\bibnamefont
  {Kraemer}}, \bibinfo {author} {\bibfnamefont {C.}~\bibnamefont {Nayak}},
  \bibinfo {author} {\bibfnamefont {M.}~\bibnamefont {Troyer}}, \bibinfo
  {author} {\bibfnamefont {C.~M.}\ \bibnamefont {Marcus}},\ and\ \bibinfo
  {author} {\bibfnamefont {C.~J.}\ \bibnamefont {Palmstr\"om}},\ }\bibfield
  {title} {\bibinfo {title} {Two-dimensional epitaxial
  superconductor-semiconductor heterostructures: A platform for topological
  superconducting networks},\ }\href
  {https://doi.org/10.1103/physrevb.93.155402} {\bibfield  {journal} {\bibinfo
  {journal} {Physical Review B}\ }\textbf {\bibinfo {volume} {93}},\ \bibinfo
  {pages} {155402} (\bibinfo {year} {2016})}\BibitemShut {NoStop}%
\bibitem [{\citenamefont {Simoni}\ \emph {et~al.}(2018)\citenamefont {Simoni},
  \citenamefont {Paolucci}, \citenamefont {Solinas}, \citenamefont
  {Strambini},\ and\ \citenamefont {Giazotto}}]{De_Simoni2018}%
  \BibitemOpen
  \bibfield  {author} {\bibinfo {author} {\bibfnamefont {G.~D.}\ \bibnamefont
  {Simoni}}, \bibinfo {author} {\bibfnamefont {F.}~\bibnamefont {Paolucci}},
  \bibinfo {author} {\bibfnamefont {P.}~\bibnamefont {Solinas}}, \bibinfo
  {author} {\bibfnamefont {E.}~\bibnamefont {Strambini}},\ and\ \bibinfo
  {author} {\bibfnamefont {F.}~\bibnamefont {Giazotto}},\ }\bibfield  {title}
  {\bibinfo {title} {Metallic supercurrent field-effect transistor},\ }\href
  {https://doi.org/10.1038/s41565-018-0190-3} {\bibfield  {journal} {\bibinfo
  {journal} {Nature Nanotechnology}\ }\textbf {\bibinfo {volume} {13}},\
  \bibinfo {pages} {802} (\bibinfo {year} {2018})}\BibitemShut {NoStop}%
\bibitem [{\citenamefont {Golokolenov}\ \emph {et~al.}(2021)\citenamefont
  {Golokolenov}, \citenamefont {Guthrie}, \citenamefont {Kafanov},
  \citenamefont {Pashkin},\ and\ \citenamefont {Tsepelin}}]{Golokolenov2020}%
  \BibitemOpen
  \bibfield  {author} {\bibinfo {author} {\bibfnamefont {I.}~\bibnamefont
  {Golokolenov}}, \bibinfo {author} {\bibfnamefont {A.}~\bibnamefont
  {Guthrie}}, \bibinfo {author} {\bibfnamefont {S.}~\bibnamefont {Kafanov}},
  \bibinfo {author} {\bibfnamefont {Y.}~\bibnamefont {Pashkin}},\ and\ \bibinfo
  {author} {\bibfnamefont {V.}~\bibnamefont {Tsepelin}},\ }\bibfield  {title}
  {\bibinfo {title} {On the origin of the controversial electrostatic field
  effect in superconductors},\ }\href
  {https://doi.org/10.1038/s41467-021-22998-0} {\bibfield  {journal} {\bibinfo
  {journal} {Nature Communications}\ }\textbf {\bibinfo {volume} {12}},\
  \bibinfo {pages} {2747} (\bibinfo {year} {2021})}\BibitemShut {NoStop}%
\bibitem [{\citenamefont {Winkler}\ \emph {et~al.}(2019)\citenamefont
  {Winkler}, \citenamefont {Antipov}, \citenamefont {van Heck}, \citenamefont
  {Soluyanov}, \citenamefont {Glazman}, \citenamefont {Wimmer},\ and\
  \citenamefont {Lutchyn}}]{Winkler2019}%
  \BibitemOpen
  \bibfield  {author} {\bibinfo {author} {\bibfnamefont {G.~W.}\ \bibnamefont
  {Winkler}}, \bibinfo {author} {\bibfnamefont {A.~E.}\ \bibnamefont
  {Antipov}}, \bibinfo {author} {\bibfnamefont {B.}~\bibnamefont {van Heck}},
  \bibinfo {author} {\bibfnamefont {A.~A.}\ \bibnamefont {Soluyanov}}, \bibinfo
  {author} {\bibfnamefont {L.~I.}\ \bibnamefont {Glazman}}, \bibinfo {author}
  {\bibfnamefont {M.}~\bibnamefont {Wimmer}},\ and\ \bibinfo {author}
  {\bibfnamefont {R.~M.}\ \bibnamefont {Lutchyn}},\ }\bibfield  {title}
  {\bibinfo {title} {Unified numerical approach to topological
  semiconductor-superconductor heterostructures},\ }\href
  {https://doi.org/10.1103/PhysRevB.99.245408} {\bibfield  {journal} {\bibinfo
  {journal} {Physical Review B}\ }\textbf {\bibinfo {volume} {99}},\ \bibinfo
  {pages} {245408} (\bibinfo {year} {2019})}\BibitemShut {NoStop}%
\bibitem [{\citenamefont {Giglberger}\ \emph {et~al.}(2007)\citenamefont
  {Giglberger}, \citenamefont {Golub}, \citenamefont {Bel'kov}, \citenamefont
  {Danilov}, \citenamefont {Schuh}, \citenamefont {Gerl}, \citenamefont
  {Rohlfing}, \citenamefont {Stahl}, \citenamefont {Wegscheider}, \citenamefont
  {Weiss}, \citenamefont {Prettl},\ and\ \citenamefont
  {Ganichev}}]{Giglberger2007}%
  \BibitemOpen
  \bibfield  {author} {\bibinfo {author} {\bibfnamefont {S.}~\bibnamefont
  {Giglberger}}, \bibinfo {author} {\bibfnamefont {L.~E.}\ \bibnamefont
  {Golub}}, \bibinfo {author} {\bibfnamefont {V.~V.}\ \bibnamefont {Bel'kov}},
  \bibinfo {author} {\bibfnamefont {S.~N.}\ \bibnamefont {Danilov}}, \bibinfo
  {author} {\bibfnamefont {D.}~\bibnamefont {Schuh}}, \bibinfo {author}
  {\bibfnamefont {C.}~\bibnamefont {Gerl}}, \bibinfo {author} {\bibfnamefont
  {F.}~\bibnamefont {Rohlfing}}, \bibinfo {author} {\bibfnamefont
  {J.}~\bibnamefont {Stahl}}, \bibinfo {author} {\bibfnamefont
  {W.}~\bibnamefont {Wegscheider}}, \bibinfo {author} {\bibfnamefont
  {D.}~\bibnamefont {Weiss}}, \bibinfo {author} {\bibfnamefont
  {W.}~\bibnamefont {Prettl}},\ and\ \bibinfo {author} {\bibfnamefont {S.~D.}\
  \bibnamefont {Ganichev}},\ }\bibfield  {title} {\bibinfo {title} {Rashba and
  dresselhaus spin splittings in semiconductor quantum wells measured by spin
  photocurrents},\ }\href {https://doi.org/10.1103/PhysRevB.75.035327}
  {\bibfield  {journal} {\bibinfo  {journal} {Physical Review B}\ }\textbf
  {\bibinfo {volume} {75}},\ \bibinfo {pages} {035327} (\bibinfo {year}
  {2007})}\BibitemShut {NoStop}%
\bibitem [{\citenamefont {Farzaneh}\ \emph {et~al.}(2022)\citenamefont
  {Farzaneh}, \citenamefont {Hatefipour}, \citenamefont {Schiela},
  \citenamefont {Lotfizadeh}, \citenamefont {Yu}, \citenamefont {Elfeky},
  \citenamefont {Strickland}, \citenamefont {Matos-Abiague},\ and\
  \citenamefont {Shabani}}]{Farzaneh2022}%
  \BibitemOpen
  \bibfield  {author} {\bibinfo {author} {\bibfnamefont {S.~M.}\ \bibnamefont
  {Farzaneh}}, \bibinfo {author} {\bibfnamefont {M.}~\bibnamefont
  {Hatefipour}}, \bibinfo {author} {\bibfnamefont {W.~F.}\ \bibnamefont
  {Schiela}}, \bibinfo {author} {\bibfnamefont {N.}~\bibnamefont {Lotfizadeh}},
  \bibinfo {author} {\bibfnamefont {P.}~\bibnamefont {Yu}}, \bibinfo {author}
  {\bibfnamefont {B.~H.}\ \bibnamefont {Elfeky}}, \bibinfo {author}
  {\bibfnamefont {W.~M.}\ \bibnamefont {Strickland}}, \bibinfo {author}
  {\bibfnamefont {A.}~\bibnamefont {Matos-Abiague}},\ and\ \bibinfo {author}
  {\bibfnamefont {J.}~\bibnamefont {Shabani}},\ }\bibfield  {title} {\bibinfo
  {title} {Magneto-anisotropic weak antilocalization in near-surface quantum
  wells},\ }\Eprint {https://arxiv.org/abs/Search...} {Search...}  (\bibinfo
  {year} {2022})\BibitemShut {NoStop}%
\bibitem [{\citenamefont {Baumgartner}\ \emph
  {et~al.}(2022{\natexlab{b}})\citenamefont {Baumgartner}, \citenamefont
  {Fuchs}, \citenamefont {Costa}, \citenamefont {Pic{\'{o}}-Cort{\'{e}}s},
  \citenamefont {Reinhardt}, \citenamefont {Gronin}, \citenamefont {Gardner},
  \citenamefont {Lindemann}, \citenamefont {Manfra}, \citenamefont {Junior},
  \citenamefont {Kochan}, \citenamefont {Fabian}, \citenamefont {Paradiso},\
  and\ \citenamefont {Strunk}}]{Baumgartner2022a}%
  \BibitemOpen
  \bibfield  {author} {\bibinfo {author} {\bibfnamefont {C.}~\bibnamefont
  {Baumgartner}}, \bibinfo {author} {\bibfnamefont {L.}~\bibnamefont {Fuchs}},
  \bibinfo {author} {\bibfnamefont {A.}~\bibnamefont {Costa}}, \bibinfo
  {author} {\bibfnamefont {J.}~\bibnamefont {Pic{\'{o}}-Cort{\'{e}}s}},
  \bibinfo {author} {\bibfnamefont {S.}~\bibnamefont {Reinhardt}}, \bibinfo
  {author} {\bibfnamefont {S.}~\bibnamefont {Gronin}}, \bibinfo {author}
  {\bibfnamefont {G.~C.}\ \bibnamefont {Gardner}}, \bibinfo {author}
  {\bibfnamefont {T.}~\bibnamefont {Lindemann}}, \bibinfo {author}
  {\bibfnamefont {M.~J.}\ \bibnamefont {Manfra}}, \bibinfo {author}
  {\bibfnamefont {P.~E.~F.}\ \bibnamefont {Junior}}, \bibinfo {author}
  {\bibfnamefont {D.}~\bibnamefont {Kochan}}, \bibinfo {author} {\bibfnamefont
  {J.}~\bibnamefont {Fabian}}, \bibinfo {author} {\bibfnamefont
  {N.}~\bibnamefont {Paradiso}},\ and\ \bibinfo {author} {\bibfnamefont
  {C.}~\bibnamefont {Strunk}},\ }\bibfield  {title} {\bibinfo {title} {Effect
  of rashba and dresselhaus spin{\textendash}orbit coupling on supercurrent
  rectification and magnetochiral anisotropy of ballistic josephson
  junctions},\ }\href {https://doi.org/10.1088/1361-648X/ac4d5e} {\bibfield
  {journal} {\bibinfo  {journal} {Journal of Physics: Condensed Matter}\
  }\textbf {\bibinfo {volume} {34}},\ \bibinfo {pages} {154005} (\bibinfo
  {year} {2022}{\natexlab{b}})}\BibitemShut {NoStop}%
\bibitem [{\citenamefont {Geim}\ \emph {et~al.}(1997)\citenamefont {Geim},
  \citenamefont {Grigorieva}, \citenamefont {Dubonos}, \citenamefont {Lok},
  \citenamefont {Maan}, \citenamefont {Filippov},\ and\ \citenamefont
  {Peeters}}]{Geim1997a}%
  \BibitemOpen
  \bibfield  {author} {\bibinfo {author} {\bibfnamefont {A.~K.}\ \bibnamefont
  {Geim}}, \bibinfo {author} {\bibfnamefont {I.~V.}\ \bibnamefont
  {Grigorieva}}, \bibinfo {author} {\bibfnamefont {S.~V.}\ \bibnamefont
  {Dubonos}}, \bibinfo {author} {\bibfnamefont {J.~G.~S.}\ \bibnamefont {Lok}},
  \bibinfo {author} {\bibfnamefont {J.~C.}\ \bibnamefont {Maan}}, \bibinfo
  {author} {\bibfnamefont {A.~E.}\ \bibnamefont {Filippov}},\ and\ \bibinfo
  {author} {\bibfnamefont {F.~M.}\ \bibnamefont {Peeters}},\ }\bibfield
  {title} {\bibinfo {title} {Phase transitions in individual sub-micrometre
  superconductors},\ }\href {https://doi.org/10.1038/36797} {\bibfield
  {journal} {\bibinfo  {journal} {Nature}\ }\textbf {\bibinfo {volume} {390}},\
  \bibinfo {pages} {259} (\bibinfo {year} {1997})}\BibitemShut {NoStop}%
\bibitem [{\citenamefont {Nichele}\ \emph {et~al.}(2017)\citenamefont
  {Nichele}, \citenamefont {Drachmann}, \citenamefont {Whiticar}, \citenamefont
  {O'Farrell}, \citenamefont {Suominen}, \citenamefont {Fornieri},
  \citenamefont {Wang}, \citenamefont {Gardner}, \citenamefont {Thomas},
  \citenamefont {Hatke}, \citenamefont {Krogstrup}, \citenamefont {Manfra},
  \citenamefont {Flensberg},\ and\ \citenamefont {Marcus}}]{Nichele2017}%
  \BibitemOpen
  \bibfield  {author} {\bibinfo {author} {\bibfnamefont {F.}~\bibnamefont
  {Nichele}}, \bibinfo {author} {\bibfnamefont {A.~C.}\ \bibnamefont
  {Drachmann}}, \bibinfo {author} {\bibfnamefont {A.~M.}\ \bibnamefont
  {Whiticar}}, \bibinfo {author} {\bibfnamefont {E.~C.}\ \bibnamefont
  {O'Farrell}}, \bibinfo {author} {\bibfnamefont {H.~J.}\ \bibnamefont
  {Suominen}}, \bibinfo {author} {\bibfnamefont {A.}~\bibnamefont {Fornieri}},
  \bibinfo {author} {\bibfnamefont {T.}~\bibnamefont {Wang}}, \bibinfo {author}
  {\bibfnamefont {G.~C.}\ \bibnamefont {Gardner}}, \bibinfo {author}
  {\bibfnamefont {C.}~\bibnamefont {Thomas}}, \bibinfo {author} {\bibfnamefont
  {A.~T.}\ \bibnamefont {Hatke}}, \bibinfo {author} {\bibfnamefont
  {P.}~\bibnamefont {Krogstrup}}, \bibinfo {author} {\bibfnamefont {M.~J.}\
  \bibnamefont {Manfra}}, \bibinfo {author} {\bibfnamefont {K.}~\bibnamefont
  {Flensberg}},\ and\ \bibinfo {author} {\bibfnamefont {C.~M.}\ \bibnamefont
  {Marcus}},\ }\bibfield  {title} {\bibinfo {title} {{Scaling of Majorana
  Zero-Bias Conductance Peaks}},\ }\href
  {https://doi.org/10.1103/PhysRevLett.119.136803} {\bibfield  {journal}
  {\bibinfo  {journal} {Physical Review Letters}\ }\textbf {\bibinfo {volume}
  {119}},\ \bibinfo {pages} {1} (\bibinfo {year} {2017})},\ \Eprint
  {https://arxiv.org/abs/1706.07033} {1706.07033} \BibitemShut {NoStop}%
\bibitem [{\citenamefont {Wei}\ \emph {et~al.}(2006)\citenamefont {Wei},
  \citenamefont {Pekker}, \citenamefont {Rogachev}, \citenamefont {Bezryadin},\
  and\ \citenamefont {Goldbart}}]{Wei2006}%
  \BibitemOpen
  \bibfield  {author} {\bibinfo {author} {\bibfnamefont {T.-C.}\ \bibnamefont
  {Wei}}, \bibinfo {author} {\bibfnamefont {D.}~\bibnamefont {Pekker}},
  \bibinfo {author} {\bibfnamefont {A.}~\bibnamefont {Rogachev}}, \bibinfo
  {author} {\bibfnamefont {A.}~\bibnamefont {Bezryadin}},\ and\ \bibinfo
  {author} {\bibfnamefont {P.~M.}\ \bibnamefont {Goldbart}},\ }\bibfield
  {title} {\bibinfo {title} {Enhancing superconductivity: Magnetic impurities
  and their quenching by magnetic fields},\ }\href
  {https://doi.org/10.1209/epl/i2006-10218-2} {\bibfield  {journal} {\bibinfo
  {journal} {Europhys. Lett.}\ }\textbf {\bibinfo {volume} {75}},\ \bibinfo
  {pages} {943} (\bibinfo {year} {2006})},\ \Eprint
  {https://arxiv.org/abs/0510476} {0510476} \BibitemShut {NoStop}%
\bibitem [{\citenamefont {Matveev}\ \emph {et~al.}(2002)\citenamefont
  {Matveev}, \citenamefont {Larkin},\ and\ \citenamefont
  {Glazman}}]{Matveev2002}%
  \BibitemOpen
  \bibfield  {author} {\bibinfo {author} {\bibfnamefont {K.~A.}\ \bibnamefont
  {Matveev}}, \bibinfo {author} {\bibfnamefont {A.~I.}\ \bibnamefont
  {Larkin}},\ and\ \bibinfo {author} {\bibfnamefont {L.~I.}\ \bibnamefont
  {Glazman}},\ }\bibfield  {title} {\bibinfo {title} {Persistent current in
  superconducting nanorings},\ }\href
  {https://doi.org/10.1103/PhysRevLett.89.096802} {\bibfield  {journal}
  {\bibinfo  {journal} {Phys. Rev. Lett.}\ }\textbf {\bibinfo {volume} {89}},\
  \bibinfo {pages} {096802} (\bibinfo {year} {2002})}\BibitemShut {NoStop}%
\end{thebibliography}




\clearpage
\newpage
\onecolumngrid

\renewcommand{\thefigure}{S\arabic{figure}}
\renewcommand{\theequation}{S\arabic{equation}}
\renewcommand{\thetable}{S\arabic{table}}
\renewcommand{\thepage}{sup-\arabic{page}}
\setcounter{page}{1}
\setcounter{equation}{0}
\setcounter{figure}{0}
\setcounter{table}{0}

\begin{center}
\textbf{\Large Diamagnetic mechanism of critical current non-reciprocity in heterostructured superconductors}\\
\textbf{\large Supplementary Materials} \\
{\it Ananthesh Sundaresh, Jukka I.~V\"ayrynen, Yuli Lyanda-Geller and Leonid P. Rokhinson}
\end{center}

\tableofcontents

\clearpage
\section{Comparison of symmetric and asymmetric contributions to the critical current.}
\begin{figure}[h]
\centering
\includegraphics[width=0.7\textwidth]{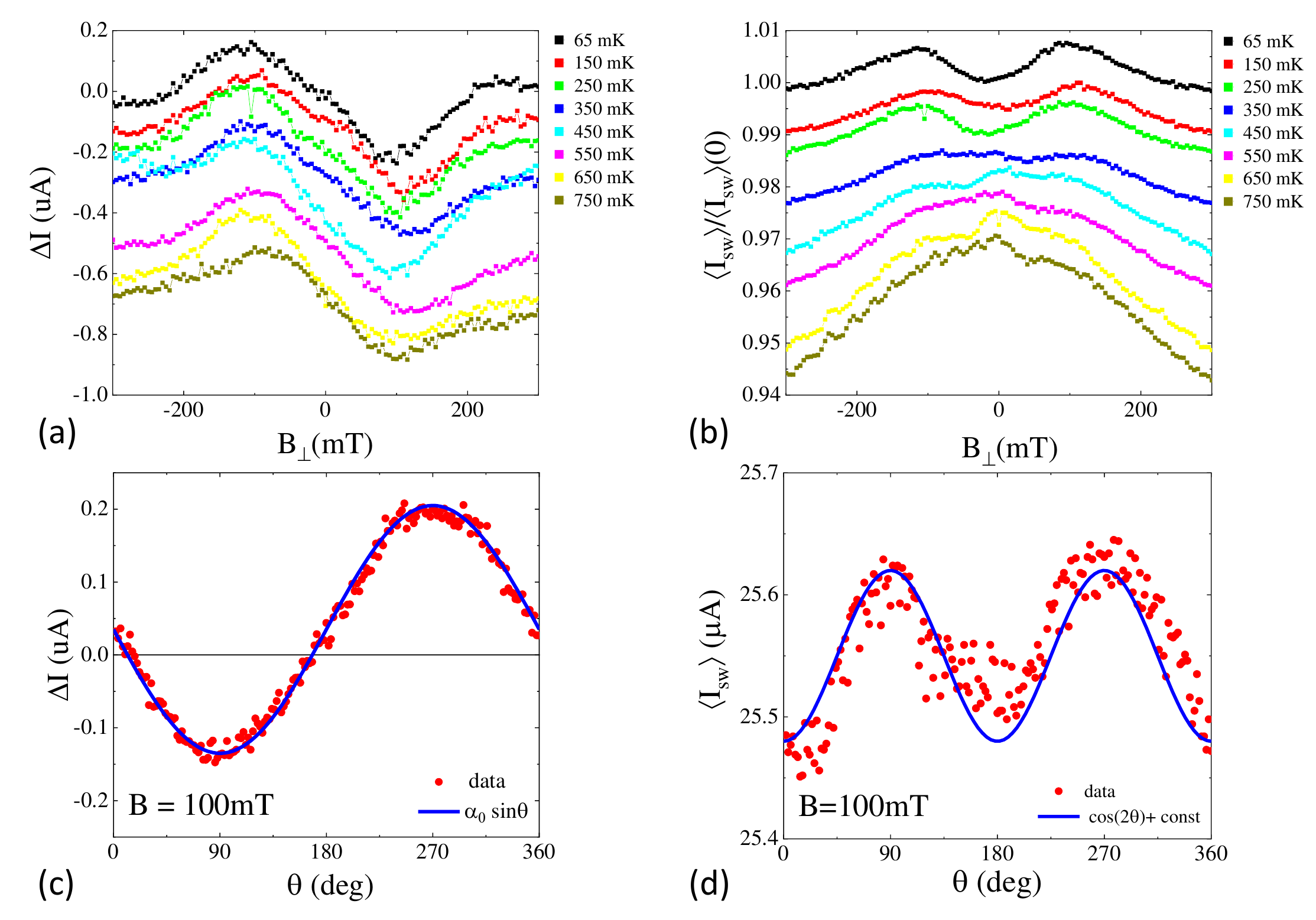}
\caption{\textbf{Field dependence of symmetric and asymmetric parts of the switching current.} (a,b) Temperature dependence of $\Delta I$ and $\langle I_{sw}\rangle$ measured as a function of $B_\bot$ ($\theta=90$ deg). The curves are offset by -0.1 $\mu$A ($\Delta I$) and 0.01 ($\langle I_{sw}\rangle$). (c,d) Angle dependence is measured by rotating magnetic field of constant magnitude $B=100$ mT at the base temperature.}
\label{sf1}
\end{figure}
The switching current can be decomposed into symmetric $\langle I_{sw}\rangle=(\langle I_{sw}^+\rangle+\langle I_{sw}^-\rangle)/2$ and asymmetric $\Delta I=\langle I_{sw}^+\rangle-\langle I_{sw}^-\rangle$ parts. Their dependence on magnetic field is plotted in Fig.~\ref{sf1}. Both $\langle I_{sw}\rangle$ and $\Delta I$ are non-monotonic functions of $B_\bot$, however they have very different $T$- and field-angle-dependencies. $\Delta I$ is almost unaffected by temperature up to $T\sim 0.6 T_c$, while a dip around $B_\bot=0$ in $\langle I_{sw}\rangle$ is developed at $T<0.3 T_c$. At constant $B=100$ mT, the angular dependences are $\Delta I\propto\sin(\theta)$, but field-dependent correction to $\langle I_{sw}\rangle$ is $\propto\cos(2\theta)$. These differences in energy scales ($T$-dependence) and angular dependencies indicate that suppression of $\langle I_{sw}\rangle$ near $B=0$ and asymmetric $\Delta I$ have different physical origins. Indeed, suppression of a critical current near $B=0$ has been reported in previous works on single-layer nanowires and was attributed to the presence of quasiparticles and/or magnetic impurities \cite{Golokolenov2020,Wei2006}, which differ from geometrical effects responsible for $\Delta I (B)$ dependence.

\clearpage
\section{Dependence of NRC on Temperature.}

\begin{figure}[h]
	\centering
    \includegraphics[width=01\textwidth]{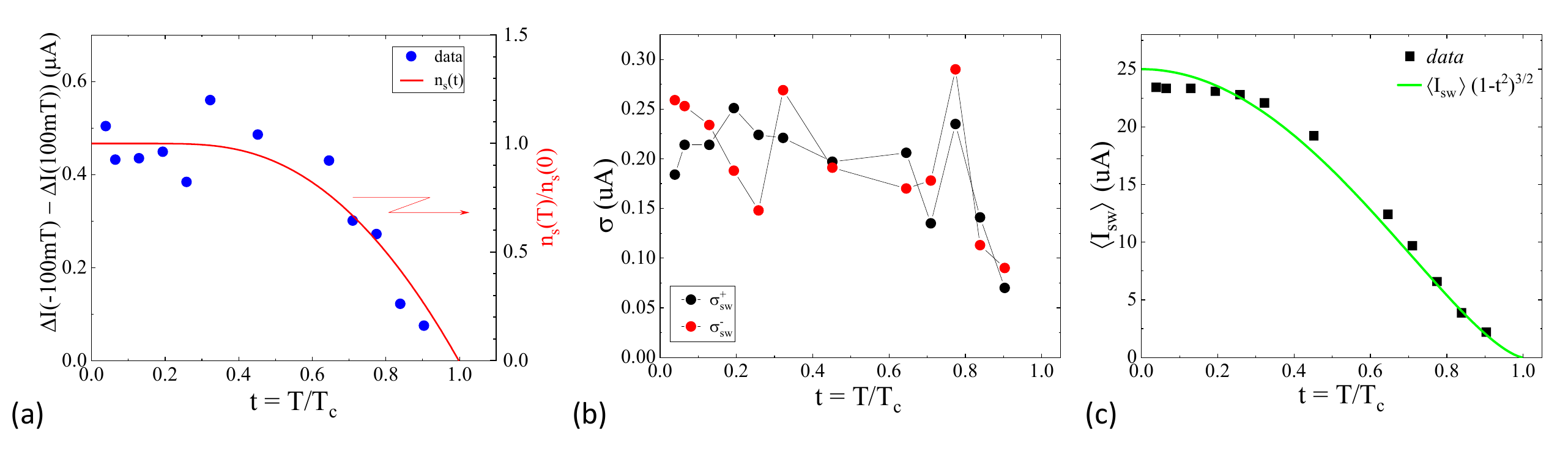}
    \caption{\textbf{The effect of temperature on NRC.} (a) The amplitude of NRC $[\Delta I(-100mT)-\Delta I(100mT)]$, (b) the standard deviation of the switching currents at \bp\ = 0, and (c) the average switching current at \bp\ = 0 are plotted as a function of the reduced temperature. NRC amplitude follows the T-dependence of the Cooper pair density $n_s(T)$, consistent with Eq.~(\ref{eq:DeltaIFinalApprox}).  $\langle I_{sw}\rangle(T)$ follows the Bardeen relation\cite{Bardeen1962}.}
\label{sf2}
\end{figure}

\clearpage
\section{Dependence of NRC on $B_\|$.}

Non-reciprocity of the switching current is linearly suppressed by an in-plane magnetic field $B_\| \| I$ and vanishes at $\approx 750$ mT. Within the same range of $B_\|$ the magnitude of the switching current remains almost constant (decreases $<2.5\%$ at $B_\|=750$ mT).

\begin{figure}[h]
	\centering
    \includegraphics[width=0.98\textwidth]{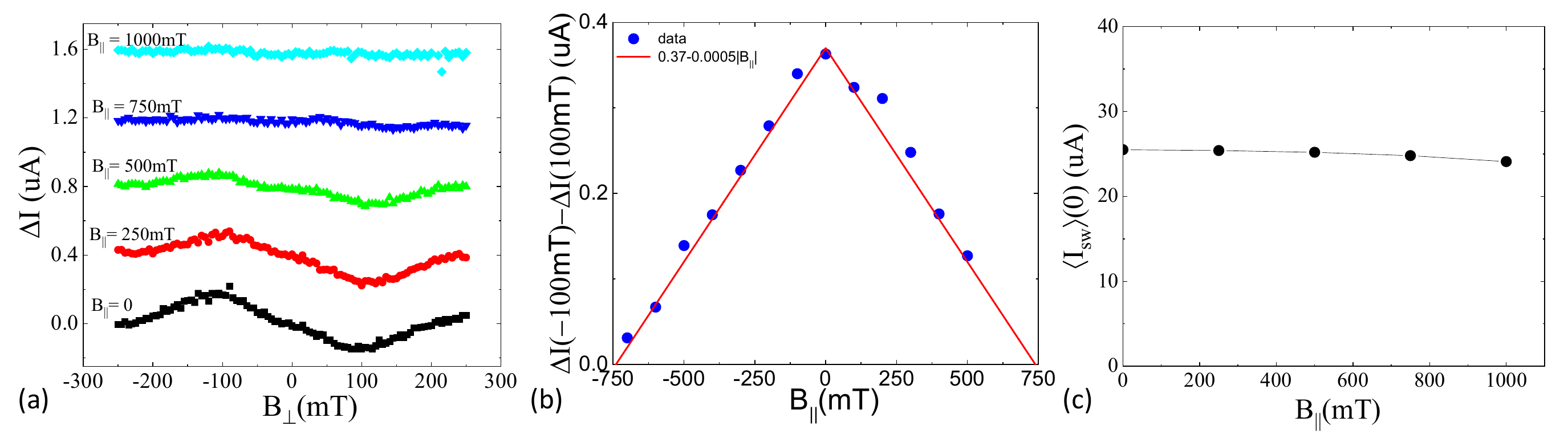}
    \caption{\textbf{The effect of an in-plane current $B||I$ on the non-reciprocal supercurrent.} (a) Evolution of \di\ in the presence of $B_{\|}$. The plots are vertically shifted for clarity (b) The NRC amplitude falls approximately linearly with $|B_\||$ (c) Dependence of average switching current \isw\ at $B_\bot=0$ on $B_\|$. All data is taken at the base temperature.}
\label{sf3}
\end{figure}

\clearpage
\section{NRC in nanowires of various width}

\begin{figure}[h]
	\centering
    \includegraphics[width=0.98\textwidth]{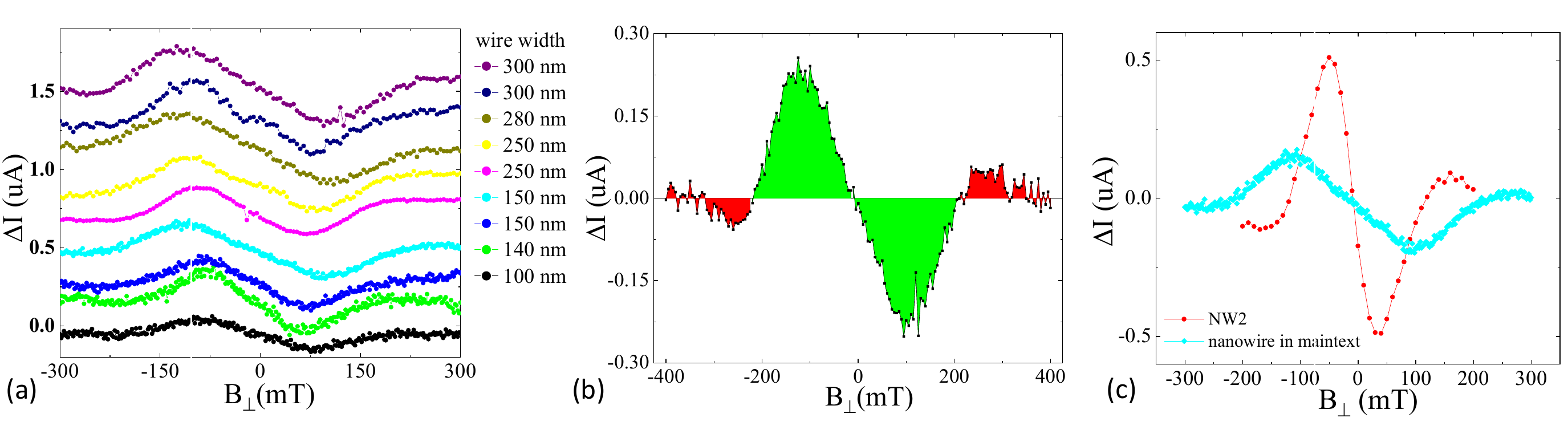}
    \caption{\textbf{NRC in other devices.} (a) NRC in nanowires of different width. The plots are shifted vertically for clarity. (b) Multiple sign reversal of \di\ in another nanowire. (c) one out of $\sim20$ nanowires fabricated from similar wafers showed enhanced magnitude of \di\ and reduced $\Delta B$.}
    \label{sf4}
\end{figure}

We have studied NRC in several nanowires of different width and length. Since all devices were fabricated from similar wafers, the Josephson coupling and, therefore, $l_J$ are similar in all devices, and we expect the amplitude of \di\ and period $\Delta B$ to be similar. Indeed, that is the case for most devices, see Fig.~\ref{sf4}a. One nanowire showed $\approx 2\times$ enhancement of \di\ and $\approx 2\times$ reduction of $\Delta B$, which would be consistent with a local enhancement of $l_J$ by a factor of 2.

\clearpage
\section{Gate dependence of NRC and critical current.}

On one of the samples we fabricated an electrostatic gate which covered the wire and a surrounding InAs 2D gas. In order to deplete electrons in InAs in the regions where it is not screened by Al, we apply a large negative gate voltage.  We see no observable effect on the NRC when varying the gate voltage. InAs is expected to be fully depleted for applied gate voltage -1.5V. We measured NRC at different gate voltages  varying from 0 to -4.5V and observed no variation of $\alpha=d\Delta I/dB_\bot$ near $B=0$ or $\Delta B$. Slight (up to 0.26\%) increase of \isw\ at large negative gate voltages is observed. Negative gate voltage also depletes carriers in Al (albeit their negligibly small fraction) and, thus, should result in the {\it decrease} of $I_c$, contrary to the observed increase. The observed increase of the switching current may result from the reduction of quantum fluctuations due to the reduction of InAs volume for Cooper pairs to enter and, as a consequence, increasing switching current to be closer to the value of the critical current.

 	\begin{figure}[h]
 		\centering
 		\includegraphics[width=.98\textwidth]{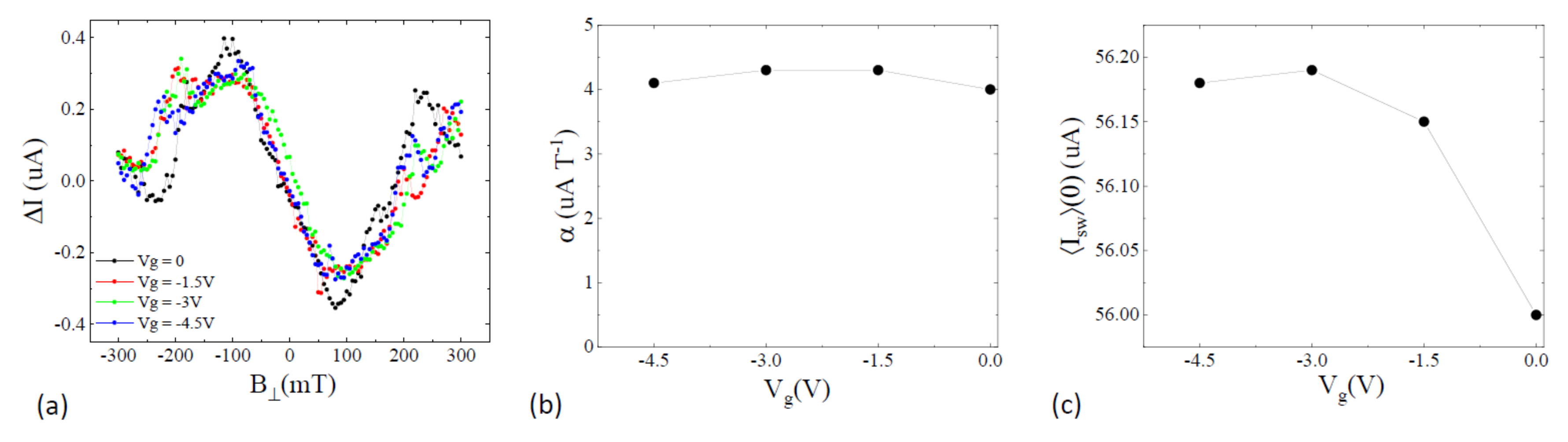}\\
 		\caption{\textbf{The effect of gate voltage on NRC.} (a) NRC shows no observable difference on varying the gate voltage. (b) $\alpha$ shows very little variation with gate voltage. (c) When a negative gate voltage is applied the \isw\ increases.
 		}
 		\label{dibpvsvg}
  	\end{figure}

\clearpage
\newpage
\section{Absence of NRC in aluminum nanowire.}

\begin{figure}[h]
	\centering
    \includegraphics[width=0.7\textwidth]{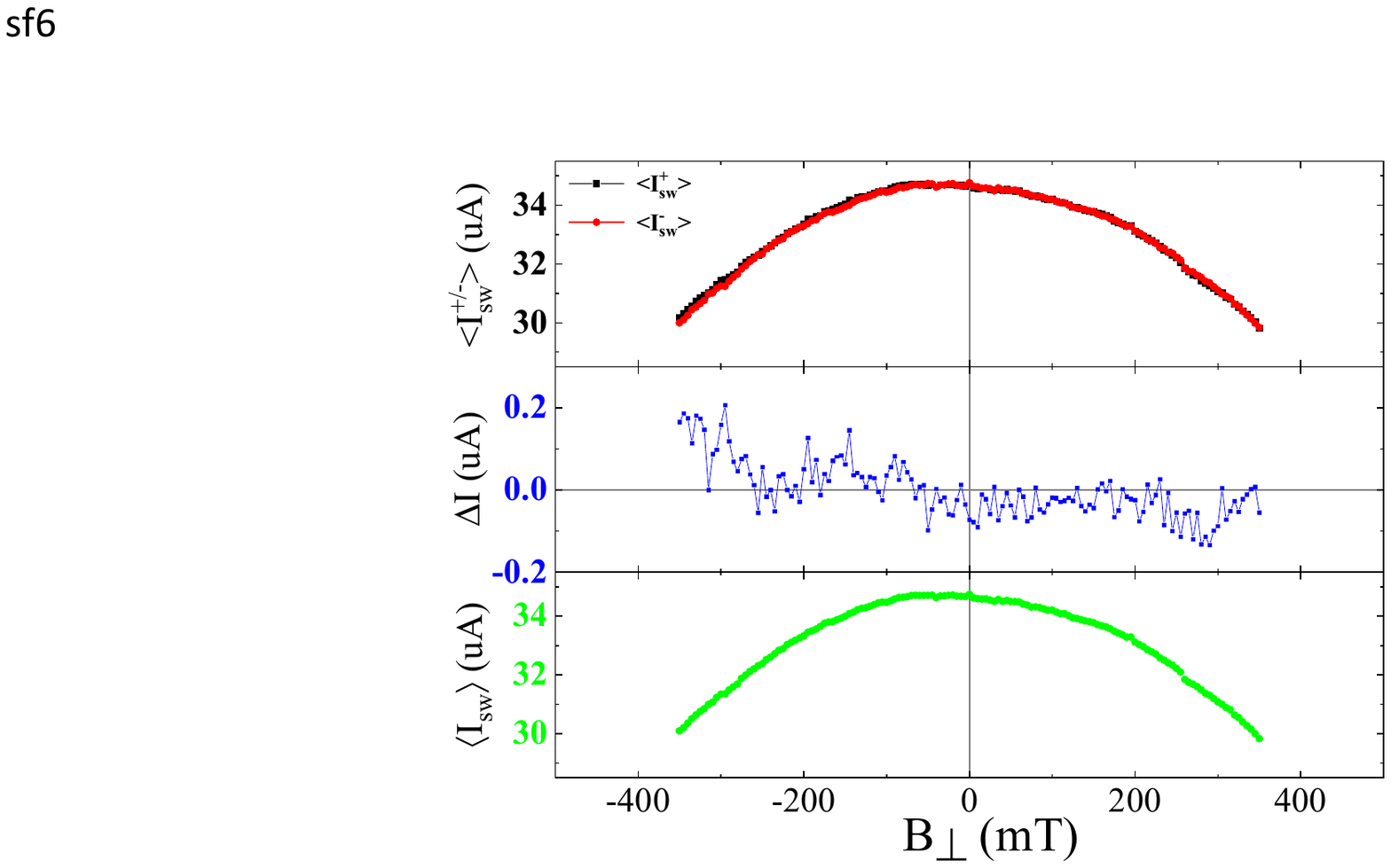}
    \caption{\textbf{No NRC in a control device.} A control 150 nm wide and 3 $\mu$m long nanowire is fabricated from a 20 nm thick Al film deposited on  a semi-insulating Si wafer. This device shows no NRC.}
\end{figure}

\clearpage
\section{In-plane magnetic field alignment}

\begin{figure}[h]
	\centering
    \includegraphics[width=0.9\textwidth]{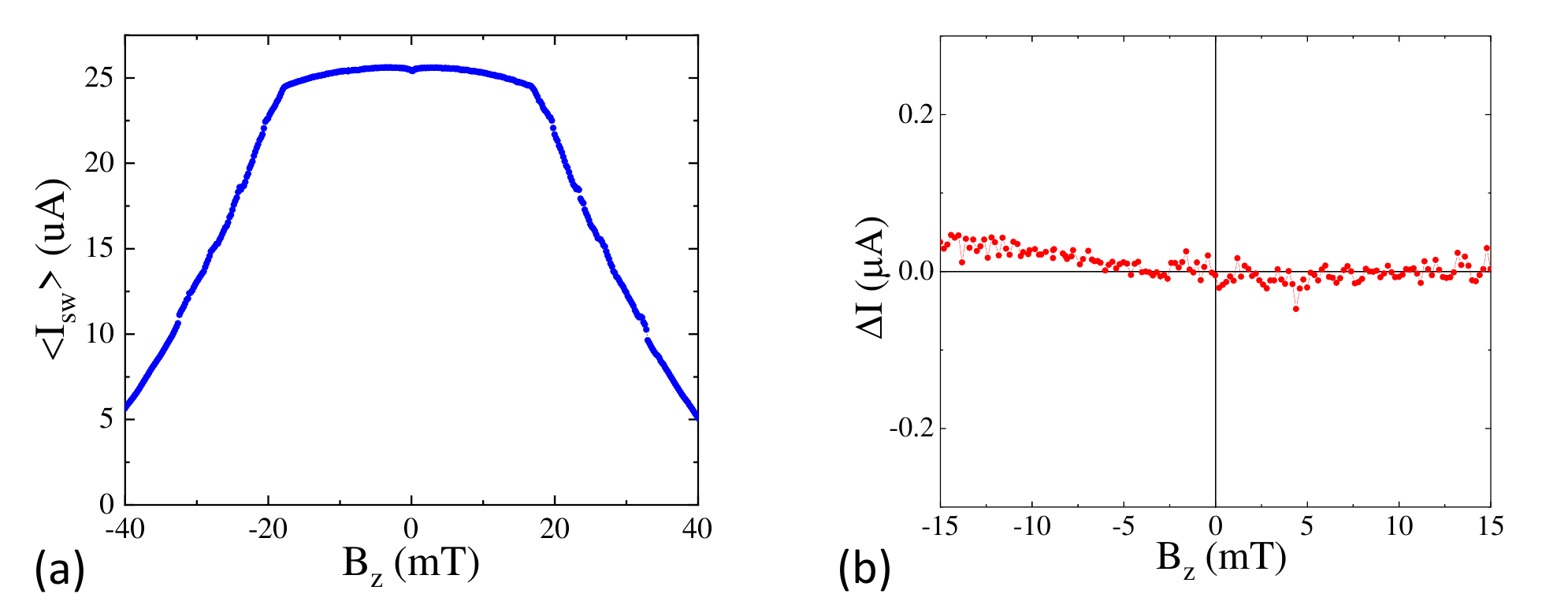}
    \caption{(a) Field dependence of \isw\ shows a Meissner state up to $B_z\approx 18$ mT. (b) No NRC is observed in an out-of- plane magnetic field.}
    \label{bzdependence}
\end{figure}

Magnetic fields are generated by a 3-axis vector magnet. The critical out-of-plane field for our wires is $B^z_{c2}\approx60$ mT. A sharp reduction of \isw\ at $B_z>18$ mT is associated with an entrance of Abrikosov vortices. In order to align the in-plane field with the plane of the sample the following alignment procedure has been used. The in-plane field was ramped to $B_\|^{'}\approx 800$ mT, beyond the field where NRC is observed. Next, $B_z$ field is scanned $\pm 30$ mT and a symmetry point $B_z^{'}$ is determined. In subsequent scans a linear correction $B_z=a B_\|$, where $a=B_z^{'}/B_\|^{'}$, is applied to keep $B_\|$ aligned with the sample plane with a precision of $<0.1$\ degree.

\clearpage
\section{Theory: Geometric effects and $I_c$ non-reciprocity in coupled superconducting wires}
\label{sec:2-wire}

\begin{figure}[b]
 		\centering
 		\includegraphics[width=\textwidth]{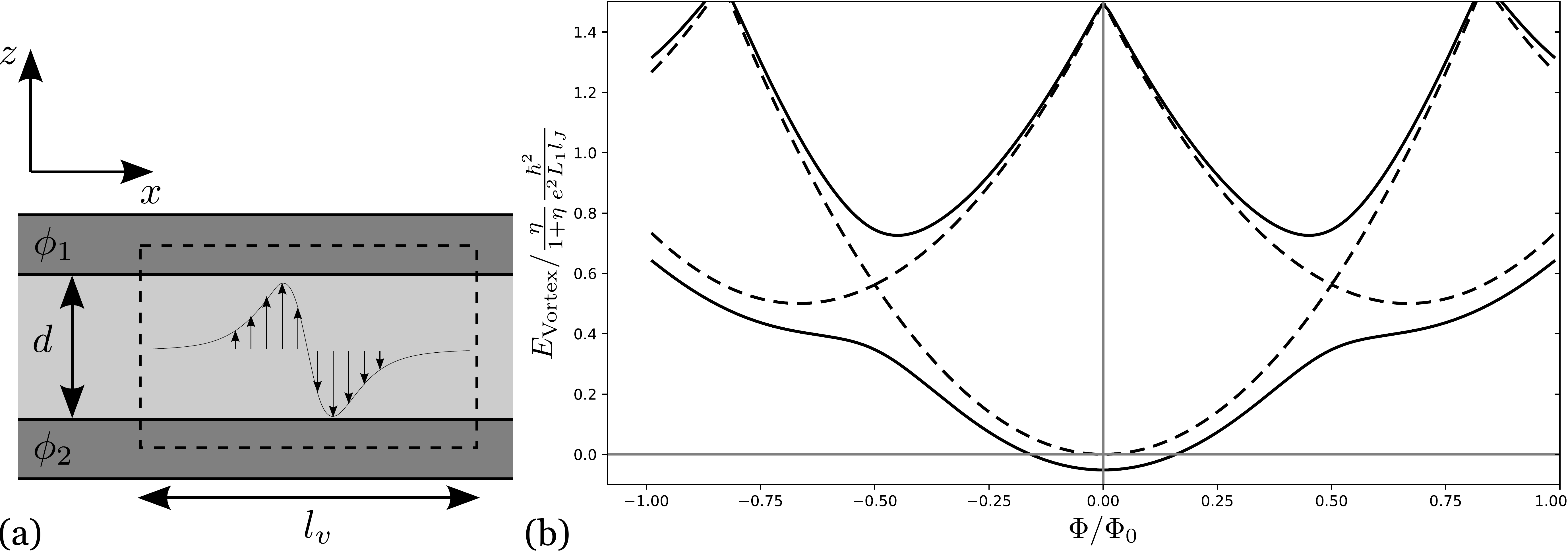}
 		\caption{
 		(a)
 		Schematic picture of the model to explain non-reciprocity. The dark grey regions depict two superconducting wires labeled 1 and 2 (corresponding to Al and proximitized InAs wires, respectively) with the order parameter phases $\phi_1$, $\phi_2$.
 		The region between the wires denotes the insulating barrier of thickness $d$.
 		In most positions $x$, the phases are locked to $\phi_{1}=\phi_{2}\,(\mathrm{mod}\,2\pi)$ due to a strong Josephson coupling.
        In the region of length $l_v$ spanned by the Josephson vortex         the phases are not equal and as a result the phase difference winds by an additional $2 \pi n$ over the vortex. The vertical arrows denote the resulting Josephson currents flowing between the two wires in the vortex.
        (b) Total energy vs magnetic field in the two-wire model.
 		The dashed curves show the spectrum obtained from Eq.~(\ref{eq:EnVx}). The three parabolas correspond to  Josephson  vortices/antivortices with $n=-1,0,1$.
 		The solid curves show the energies when coherent  vortex tunneling (strength $E_t = 0.4 $ in units of $\frac{\eta}{1+\eta}\frac{ \hbar^{2} }{e^2  L_{1} l_{J}}$) is included, leading to avoided crossings of states with different $n$.
 		}
 		\label{fig:ringSM}
  	\end{figure}

In this section we derive the critical superconducting current in two Josephson-coupled superconducting wires (the ``two-wire model'') in the presence of an external magnetic field. We show that at high enough magnetic field, it becomes energetically favorable to form a Josephson vortex (Fig.~\ref{fig:ringSM}a), which in turn can lead to an oscillatory non-reciprocity of the critical current (Fig.~\ref{fig:ring} of the main text).
Furthermore, the oscillations will be damped due to one of the wires turning normal upon increasing the magnetic field.

Let us consider a pair of parallel superconducting wires 1 (Al wire)
and 2 (proximitized InAs) along the $x$-direction with a magnetic
field $B_{\perp}$ in the $y$-direction, normal to the plane containing two wires.
The corresponding vector potential is $A_{x}(z)=B_{\perp} z$ with the
two wires separated by distance $d$ being at positions $z=z_{1,2}=-(-1)^{1,2}d/2$,
see Fig.~\ref{fig:ringSM}a. We ignore here the screening of  the magnetic field by the Josephson vortex; this effect would merely modify the Josephson length $l_J$ (introduced below). We also approximate the wires as one-dimensional, given that their typical thickness is smaller than the penetration depth and the width is smaller that the size of the Peal vortex. This makes the supercurrent distribution approximately uniform within the wire. Denoting $\phi_{1,2}$ the phases of the superconducting
order parameters, we have a supercurrent in wire $i$ given by
\begin{equation}
I_{i}=\frac{1}{2e L_{i}}\left(\hbar\partial_{x}\phi_{i}-2eA_{x}(x, z_{i})\right)\,,\label{eq:Supercurrent2wire}
\end{equation}
in terms of the  kinetic inductances (per length) $L_i = m_{i}/(e^{2} S_{i}n_{i})$ for wires $i=1,2$. Here $S_{i}$, $m_{i}$, and $n_{i}$ denote the cross-sectional area, the effective mass, and the Cooper pair densities.
 For Al wire ($i=1$) we will account for disorder by multiplying $n_{i}$  by $\sqrt{l/\xi}$, where $l \approx 2 \mathrm{nm}$ is the mean free path and $\xi \approx 1 \mu\mathrm{m}$ is the coherence length~\cite{Bardeen1962}.
Thus, we use $L_1 \to L_1 \sqrt{\xi / l}$ in our final estimates.

The phases $\phi_{1,2}(x)$ can be found by minimizing the total energy
\begin{equation}
E_{\text{tot}}=\int dx \left[\frac{1}{2}L_{1}I_{1}^{2}+\frac{1}{2}L_{2}I_{2}^{2}  -\mathcal{E}_{J}\cos(\phi_{1}-\phi_{2})\right] \,, \label{eq:En_tot}
\end{equation}
that includes kinetic energies of each wire and a Josephson energy density $\mathcal{E}_{J}$ coupling the two wires.
In the presence of an applied external supercurrent $I_{ext}$, there is a constraint  $I_{1}(x)+I_{2}(x)=I_{ext}$ at every point $x$.
The constrained energy minimization leads to the Sine-Gordon equation for $\varphi = \phi_{1}-\phi_{2}$,
\begin{equation}
    \frac{\partial^{2}\varphi}{\partial x^{2}} =   l_J^{-2}\sin\varphi \,, \label{eq:SG}
\end{equation}
where $l_J= 1 /\sqrt{8 e^2 \mathcal{E}_{J}(L_1 + L_2) / \hbar^2 }$ is the Josephson length that determines the characteristic size of a Josephson vortex. We now solve Eq.~(\ref{eq:SG}) with the appropriate boundary conditions. We assume that the Josephson coupling in Eq.~(\ref{eq:En_tot}) is strong, such that $\phi_{1}=\phi_{2}\,(\mathrm{mod}\,2\pi)$ for most $x$. If the two phases were locked for \emph{all} $x$, i.e. $\varphi(x) = 0 \,(\mathrm{mod}\,2\pi)$, we would find a non-reciprocal critical
current $I_{c}(B_{\perp})$ with the non-reciprocity $\Delta I=I_{c,+}-I_{c,-}$
that increases monotonically with $B_{\perp}$. Experimentally, a non-monotonic
dependence is observed, see Fig.~\ref{f1}b.

The non-monotonic $\Delta I$ can be explained by a formation of a Josephson vortex, see Fig.~\ref{fig:ring}. In the Josephson vortex, the phase difference $\varphi$ increases by $2\pi$ approximately over the distance $2 \pi  l_J$; explicitly, $\varphi(x)=4\arctan e^{ x/ l_J}$ for a vortex at $x=0$.

The Josephson vortex solution yields a current distribution
\begin{flalign}
I_{1}(x) & =\frac{1}{1+\eta}I_{ext}+\delta I_n(x)\,,\\
I_{2}(x) & =\frac{\eta}{1+\eta}I_{ext}-\delta I_n(x)\,,\\
\delta I_n(x) & = \frac{2\eta}{1+\eta}\frac{1}{2e  L_{1}}\frac{\hbar}{l_J}\left(n\, \text{sech}\frac{x}{l_J}- \frac{3 }{\pi} \frac{\Phi}{\Phi_{0}}\right)\,, \label{eq:deltaIn}
\end{flalign}
where we introduced an integer index $n$, $n=\pm 1$ for the Josephson vortex/antivortex and $n=0$ in the absence of the vortex.
The vortex is centered at $x=0$, which also turns out to be the position of the maximal circulating currents in the wires 1 and 2. We denote $\eta =L_1 / L_2 =S_{2}\frac{n_{2}}{m_{2}}/S_{1}\frac{n_{1}}{m_{1}}$ and introduce the flux $\Phi/\Phi_{0}= S_v  B_{\perp} /(\pi \hbar/e)$ through the effective vortex area $S_v= (\pi^2/3) l_Jd$.

The formation of the Josephson vortex becomes energetically favorable  at a large enough magnetic field $B_\perp$.
The energy cost is determined from Eq.~(\ref{eq:En_tot}) by the balance of the Josephson energy $E_J$ lost and the kinetic energy gained in the creation of a vortex. Ignoring $n$-independent terms, we find (see Fig.~\ref{fig:ringSM}b),
\begin{equation}
    E_{\text{Vortex}}(n)
    = \frac{\eta}{1+\eta}\frac{ \hbar^{2} }{e^2  L_{1} l_{J}}
    \left[     \left(n-\frac{3}{2}\frac{\Phi}{\Phi_{0}}\right)^{2} + \frac{1}{2}|n| \right]
    \label{eq:EnVx}
\end{equation}
where $n=0,\pm1$. This energy is analogous to the (inductive) energy of a superconducting ring with a phase winding $2 \pi n$~\cite{Matveev2002} apart from the last term in Eq.~(\ref{eq:EnVx}) which is the cost in Josephson energy.
In the absence of quantum fluctuations and at $T=0$, one finds from  Eq.~(\ref{eq:EnVx}) that the thermal average $\langle n \rangle = [ \Phi/\Phi_{0}]$ is given by the nearest integer to $  \Phi/\Phi_{0}$, leading to a sawtooth-like dependence for $\Delta I$ versus $B_\perp$ (see below).
Fluctuations will smear out the sawtooth dependence. In analogy to a superconducting ring~\cite{Matveev2002}, we expect to find a harmonic dependence on the flux on a linear background in the case of strong  quantum or thermal fluctuations,
\begin{equation}
\langle n \rangle =   \frac{\Phi}{\Phi_{0}}   -\delta n \sin\frac{2\pi   \Phi}{\Phi_{0}} \,,
 \label{eq:average_nApprox}
\end{equation}
where $\delta n \ll 1$ due to strong fluctuations.
Importantly, in the case of quantum fluctuations, $\delta n$ is independent of the temperature, whereas for thermal fluctuations one has exponential dependence on $1/T$. As we discuss below, the harmonic dependence on the flux translates to a similar  dependence in the non-reciprocal part $\Delta I$ of the critical current, in agreement with experimental data. The observed weak $T$-dependence in  Fig.~\ref{sf1}a  indicates that quantum fluctuations exceed thermal fluctuations in the experiment.

\begin{figure}[b]
 		\centering
 		\includegraphics[width=.99\textwidth]{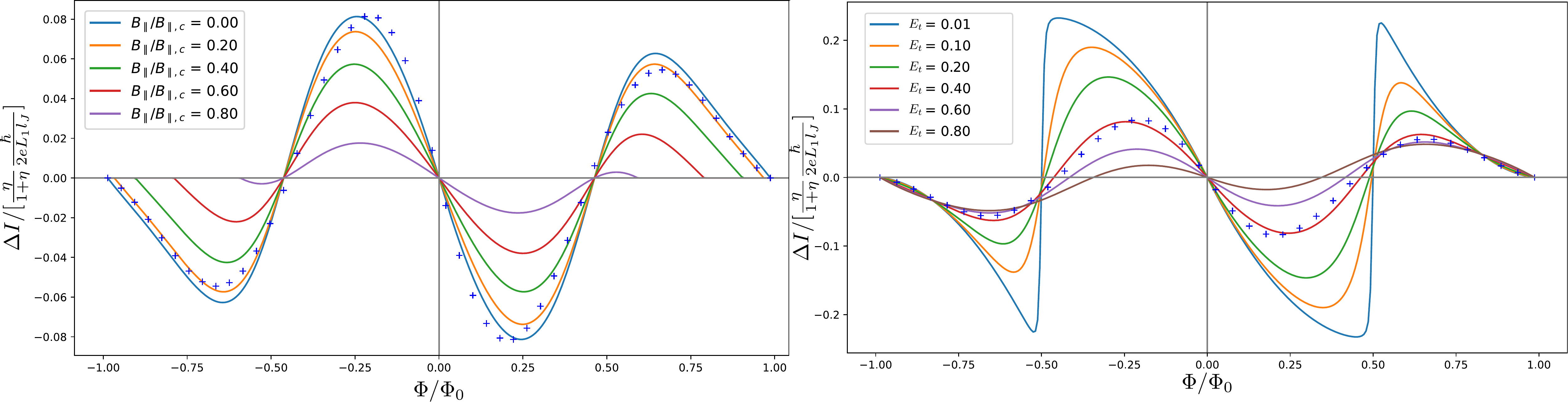}
 		\caption{
 		\textit{Left: }
 	The critical current non-reciprocity $\Delta I$, Eq.~(\ref{eq:DeltaIFinal}),  versus the flux $\Phi =  B_\perp d l_v$ through the Josephson vortex. The crosses correspond to the approximation, Eq.~(\ref{eq:DeltaIFinalApprox}).
        The applied field $B_\parallel$ suppresses the proximity effect and therefore $\Delta I$.
        In the figure $E_t = 0.4$ in units of $\frac{\eta}{1+\eta}\frac{ \hbar^{2} }{e^2  L_{1} l_{J}}$.
        \textit{Right:} $\Delta I$ (at $B_\parallel = 0$) for different strengths $E_t$ (in units of $\frac{\eta}{1+\eta}\frac{ \hbar^{2} }{e^2  L_{1} l_{J}}$) of coherent vortex tunneling that controls vortex number fluctuations; weak tunneling leads to a sawtooth-like $\Delta I$.
 		}
 		\label{fig:SMDeltaI}
  	\end{figure}

The critical current through our two wire system with contacts, effectively forming a ring-like structure  is determined by the condition that at large enough $I_{ext}$, one of the wires ( arms of the ring) turns normal.
(Experiment indicates that the switching happens in Al, i.e., wire 1, see below.)
 Assuming that $I_{ext}>0$
and the wire 1 turns normal, the corresponding condition is $I_{ext}=I_{c,+}$,
where
\begin{equation}
I_{c,\pm}=(1+\eta)(\pm I_{1,c}-\delta I)\,, \label{eq:IcpmTheory}
\end{equation}
$I_{1,c}$ is the critical current of wire 1 and $\delta I = \langle \delta I_n(0) \rangle $ is the circulating current at its peak value at $x=0$.
Likewise, for
$I_{ext}<0$ we find $I_{ext}=I_{c,-}$. This yields
\begin{flalign}
\Delta I & =I_{c,+}+I_{c,-}\\
  &
  =  -2\eta \frac{1}{L_{1}e}\frac{\hbar}{l_J}\left( \langle n \rangle \,  - \frac{3}{\pi}  \frac{\Phi}{\Phi_{0}}\right)
 \,, \label{eq:DeltaIAvApprox}
\end{flalign}
which determines  the slope $\alpha=d\Delta I/dB_{\perp}$. We note that if the wire 2 is normal, then $n_{2}=0$, $\eta=0$ and $\Delta I$ vanishes.
The $B_{\perp}$-dependence of the critical current $I_{c,\pm}$ in such case would merely show a monotonic decrease (without non-reciprocity) corrresponding to suppression of the superconducting gap in Al.
Experiments show a few distinct oscillations in the asymmetric non-reciprocal part $\Delta I$ of the critical current, see Figs.~\ref{f1}c and~\ref{sf3}.
We attribute the experimentally observed vanishing amplitude of $\Delta I$ (loss of non-reciprocity) at fields higher than $B_{\perp} \approx 750 \mathrm{mT}$ to the destruction of proximity effect. We can model this by taking $\eta$ in Eq.~(\ref{eq:DeltaIAvApprox}) to be magnetic field -dependent, detailed below.

Proximity effect is also destroyed by an in-plane field $B_{\parallel}$ along the wire (along $x$) at roughly the same $750 \mathrm{mT}$ scale, see Fig.~\ref{sf2}a.
Since the wire 2 is proximitized in our model, we include a linear in the field suppression of the Cooper pair density $n_{2}$
 at fields lower than those describing the superconducting gap suppression in the Al wire.
This leads to $\eta=\eta_{0}{(1-|\mathbf{B}|/B_{\mathrm{InAs,c}})}$ in Eq.~\ref{eq:DeltaIAvApprox}.  Here We take $B_{\mathrm{InAs,c}} \approx 750 \mathrm{mT}$ and denote $|\mathbf{B}| = \sqrt{B_{\perp}^2 + B_{\parallel}^2 }$ assuming that the suppression of proximity is isotropic (in a magnetic field parallel to heterostructure layers).
The linear suppression is taken to match with experimental observations.
In particular, a linear field-dependence is seen in Fig.~\ref{sf3}b where the slope $\alpha$ is plotted as a function
of $B_{\parallel}$. The  measurement shows also that the switching current does not differ much from its $B_{\perp}=0$ value (see Fig.~\ref{sf2}c), indicating that the critical current is determined by Al wire, as we assumed in Eq.~(\ref{eq:IcpmTheory}).

We thus obtain the following expression for the non-reciprocal contribution to the critical current, plotted in  Fig.~\ref{fig:ring} and Fig.~\ref{fig:SMDeltaI},
\begin{flalign}
    \Delta I(B_{\perp},B_{\parallel})
     & =
     -2\eta_0  \frac{1}{L_{1}e}\frac{\hbar}{l_J}\left( \langle n \rangle \, -\frac{3}{\pi}  \frac{\Phi}{\Phi_{0}}\right)
     \left(1-\frac{|\mathbf{B}|}{B_{\mathrm{InAs,c}} }\right)
     \label{eq:DeltaIFinal}
     \\
     & \approx
     -2\eta_0  \frac{1}{L_{1}e}\frac{\hbar}{l_J}\left( c \frac{\Phi}{\Phi_{0}}
     -\delta n \sin\frac{2\pi  \Phi}{\Phi_{0}}  \right)
     \left(1-\frac{|\mathbf{B}|}{B_{\mathrm{InAs,c}} }\right)\,,
    \label{eq:DeltaIFinalApprox}
\end{flalign}
where $c=  ( 1- \frac{3}{\pi} )  \approx 0.05 $ and we assumed strong quantum fluctuations of $n$,  see discussion below Eq.~(\ref{eq:average_nApprox}).
The approximate period is $ B_\perp  = \Phi_0 (3 / \pi^2) / (l_J d)  $, experimentally observed to be approximately $ 400\mathrm{mT}$. This period indicates $ 500 \mathrm{nm}$ for the effective size of the vortex, given that $d = 10\mathrm{nm}$.

From Eq.~(\ref{eq:DeltaIFinalApprox}) we obtain a zero-field slope $d \Delta I / d B_\perp \approx  -c_0 \eta_0  d / L_1  $
where $c_0 = (2\pi /3)  \left( c
     -\delta n 2\pi     \right)  $ is an unknown numerical coefficient (since $\delta n$ is unknown).
However, the dimensionless quantity $\delta n_q \ll 1$ characterizes the amplitude of the persistent current in the loop (Fig.~\ref{fig:ring}) and is suppressed due to quantum phase slips~\cite{Matveev2002}. We can therefore take $c_0 \approx  2 \pi c/3 \approx 0.1$.
Using values $S_{1}=150\mathrm{nm}\times10 \mathrm{nm}$,   $n_{1}=18\cdot10^{28}\text{m}^{-3}$ and $m_{1}=9.1\cdot10^{-31}\text{kg}$ (Al electron density and effective mass), we obtain $d/L_1 =
\sqrt{l/\xi} S_{1}\frac{e^2 n_{1} d}{m_{1}} \approx 3.4  \mathrm{mA}/\mathrm{T}$. By comparing to the  zero-field slope $d \Delta I / d B_\perp \approx 1.6 \mu \mathrm{A}/\mathrm{T} $  in Fig.~\ref{sf2}b, we obtain $\eta_0  \approx 10^{-2}$. This is consistent with an estimate $\eta_0 \approx 10^{-2}$ based on the ratio of Al and InAs kinetic inductances. We note that non-reciprocal component is proportional to the Cooper pair density,  $\Delta I \propto L_2^{-1} \propto n_2$, which is consistent with the temperature-dependence of both quantities plotted in Fig.~\ref{sf2}a.

Different properties of wires 1 and 2, i.e., their asymmetry, is essential to get non-reciprocity in our model. If the wires were identical, the wire that switches to normal state first [in  Eq.~(\ref{eq:IcpmTheory})] would change upon reversing the current direction. We note that non-reciprocity emerges even if there is no loop ($l_J \to 0$) due to Josephson vortex and no phase winding, $n=0$ from Eq.~(\ref{eq:EnVx}), but there is nevertheless a circulating  diamagnetic current $I_{dia}$, Eq.~(\ref{eq:deltaIn}), leading to non-reciprocity, Eq.~(\ref{eq:DeltaIFinal}), due to the assumed Josephson coupling induced phase locking $\phi_1 = \phi_2$ between the wires.

\clearpage
\section{Determining total critical current in a 2 wire model}
\begin{figure}[h]
 		\centering
 		\includegraphics[width=\textwidth]{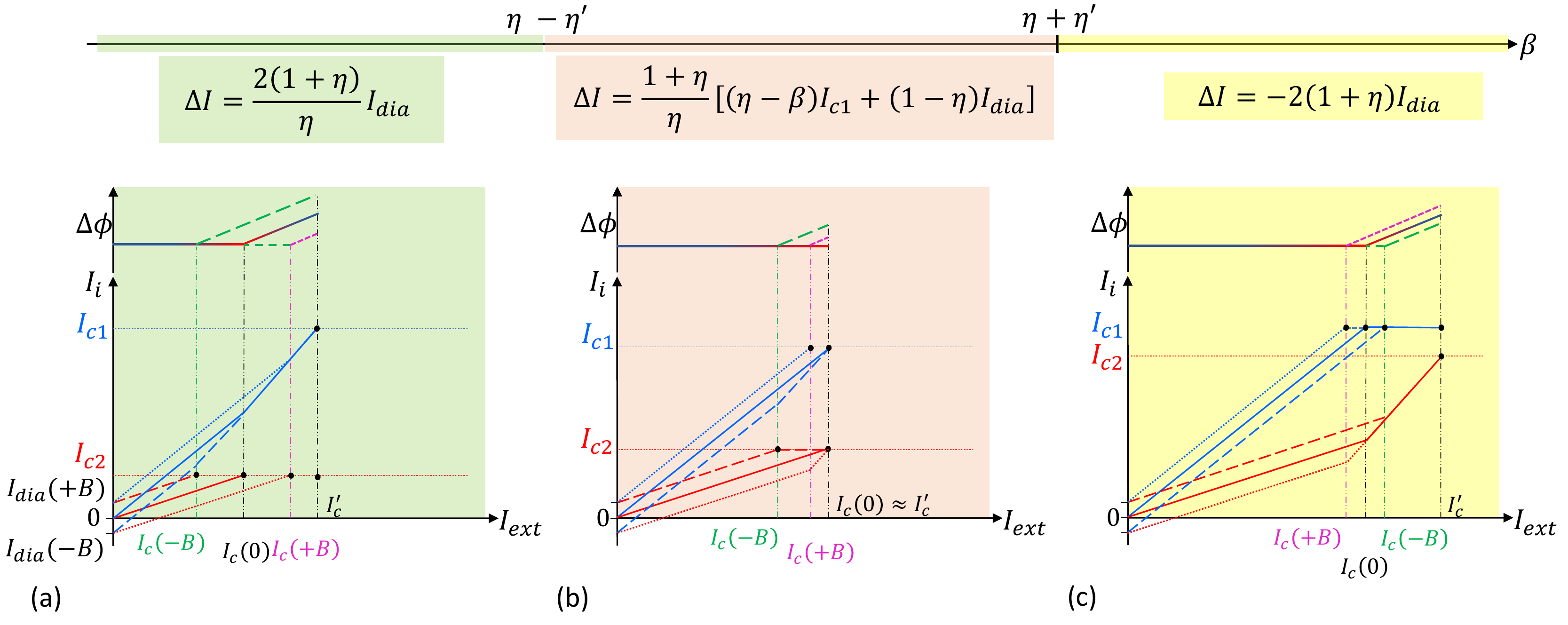}
 		\caption{Schematic of current distribution between the wires $I_i=I_{0i}-(-1)^i I_{dia}$, $i=1, 2$, and the phase difference $\Delta\phi$ as a function of an external current $I_{ext}=I_1+I_2$ for (a) $\beta\leq\eta -\eta'$, (b)$\eta -\eta'<\beta<\eta +\eta'$  and (c) $\beta\geq\eta +\eta'$. }
 		\label{sf10}
\end{figure}

The total critical current in a two-wire model depends on the two dimensionless parameters: the ratio of kinetic inductances $\eta=L_{k1}/L_{k2}$  and the ratio of critical currents $\beta=I_{c2}/I_{c1}$.
(The former also determines the current distribution in the absence of magnetic field, $\eta = I_2/I_1$.)
In a magnetic field, the total critical current will also depends on the diamagnetic current, which enters via the dimensionless ratio $\eta^{\prime}=(\eta+1)I_{dia}/I_{c1}$.
The expression of NRC depends on the magnitude of  $\beta\gtrless\eta+\eta^{\prime}$.

For $\beta\geq\eta + \eta'$, the wire 1 turns normal first for both directions of $B_y$, $I_{c}(B) = (1+\eta)(I_{c1} - I_{dia})$ and
\begin{equation}
    \Delta I = -2(1+\eta)I_{dia}
    \label{eq:case1}
\end{equation}

For $\beta\leq\eta - \eta'$, the wire 2 turns normal first for both directions of $B_y$ and
\begin{equation}
    \Delta I = +2\frac{1+\eta}{\eta}I_{dia}
\end{equation}

Finally, for $\eta - \eta'\leq\beta\leq\eta + \eta'$, the wire 1 turns normal first for $B_y>0$ and the wire 2 turns normal first for $B_y<0$ resulting in
\begin{equation}
    \Delta I = \frac{1+\eta}{\eta}\left[(\eta - \beta)I_{c1} + (1-\eta)I_{dia}\right]
\end{equation}
The current distribution between the wires for these three cases is shown schematically in Fig.~\ref{sf10}

We note that the slope $d \Delta I/ d B_{\perp}$ differs by a large factor $1/\eta$ depending on which wire turns normal first. Our data is consistent with Al (wire 1) turning normal first (at that point the whole structure is turned normal).

\end{document}